\documentclass[superscriptaddress, onecolumn, amsmath,amssymb, aps, pra, longbibliography]{revtex4-2}

\usepackage[english]{babel}
\usepackage{amsmath,amssymb}
\usepackage{physics}
\usepackage{graphicx,xcolor}
\usepackage[colorlinks=true, allcolors=blue]{hyperref}
\usepackage{ulem}
\usepackage{subcaption}
\usepackage{mathtools}
\usepackage{natbib}
\usepackage{chngcntr}
\usepackage{comment}
\usepackage{amsthm,thmtools,thm-restate}
\usepackage{placeins, afterpage}

\newtheorem{thm}{Theorem}
\newtheorem{prp}{Proposition}

\begin{document}

\title{Coherence-enhanced single-qubit thermometry out of equilibrium}

\author{Gonçalo Frazao}
\email{goncalo.frazao@tecnico.ulisboa.pt}
\affiliation{Instituto Superior Técnico, Universidade de Lisboa, Portugal.}
\affiliation{Instituto de Telecomunicações, Lisbon, Portugal.}

\author{Marco Pezzutto}
\email{marco.pezzutto@ino.cnr.it}
\affiliation{Instituto Superior Técnico, Universidade de Lisboa, Portugal.}
\affiliation{Physics of Information and Quantum Technologies Group, Centro de Física e Engenharia de Materiais Avançados (CeFEMA), Portugal.}
\affiliation{PQI -- Portuguese Quantum Institute, Portugal.}
\affiliation{Istituto Nazionale di Ottica del Consiglio Nazionale delle Ricerche (CNR-INO), Largo Enrico Fermi 6, I-50125 Firenze, Italy.}

\author{Yasser Omar}
\affiliation{Instituto Superior Técnico, Universidade de Lisboa, Portugal.}
\affiliation{Physics of Information and Quantum Technologies Group, Centro de Física e Engenharia de Materiais Avançados (CeFEMA), Portugal.}
\affiliation{PQI -- Portuguese Quantum Institute, Portugal.}

\author{Emmanuel Zambrini Cruzeiro}
\affiliation{Instituto de Telecomunicações, Lisbon, Portugal.}

\author{Stefano Gherardini}
\email{stefano.gherardini@ino.cnr.it}
\affiliation{Istituto Nazionale di Ottica del Consiglio Nazionale delle Ricerche (CNR-INO), Largo Enrico Fermi 6, I-50125 Firenze, Italy.}
\affiliation{European Laboratory for Non-linear Spectroscopy, Università di Firenze, I-50019 Sesto Fiorentino, Italy.}

\begin{abstract}
The metrological limits of thermometry operated in nonequilibrium dynamical regimes are analyzed. We consider a finite-dimensional quantum system, employed as a quantum thermometer, in contact with a thermal bath inducing Markovian thermalization dynamics. The quantum thermometer is initialized in a generic quantum state, possibly including quantum coherence w.r.t.~the Hamiltonian basis. We prove that the sensitivity of the thermometer, quantified by the quantum Fisher information, is enhanced by the quantum coherence in its initial state. We analytically show this in the specific case of qubit thermometers for which the maximization of the quantum Fisher information occurs at a finite time during the transient of the thermalization dynamics. Such a finite-time sensitivity enhancement can be better than the sensitivity that is achieved asymptotically.
\end{abstract}

\date{\today}
\maketitle

\section{Introduction}

Quantum thermometry aims at inferring the temperature of a thermal bath, or thermal reservoir, through the coupling with a quantum system~\cite{MahboudiJPA2019}.

In the quantum regime, any measurement device (thus, even a thermometer) is invasive to a given extent~\cite{SevesoPRA2018,AlbarelliPRA2023}. Hence, from an estimation perspective~\cite{ParisIJQI2009,DegenRMP2017}, the ultimate goal of thermometry is to determine the conditions under which an accurate reconstruction of the temperature of a thermal bath can be effectively attained. Quantum metrology gives us the tools to achieve this task~\cite{NicholsPRA2016,SevesoPRA2017,PezzeRMP2018,ChuPRL2022}, in terms of the {\it quantum Fisher information}~\cite{Helstrom1969,Holevo1982,HyllusPRA2012,MuellerPRA2016,LiuJPAMT2020} applied to quantum thermometry~\cite{CorreaPRL2015,JorgensenPRR2020,GebbiaPRA2020}.
In the case only few measurement records can be obtained or no prior knowledge about the thermalization dynamics are available, global thermometry has been recently proposed~\cite{RubioPRL2021}. The merit of such an approach is to identify in the mean logarithmic error an appropriate figure of merit for quantum thermometry. So far, some works have already analyzed how quantum thermometry can be employed in several quantum platforms for quantum technology. Among them, we would mention a three-level transmon circuit~\cite{SultanovAPL2021}, a pair of trapped ions~\cite{SaNetoSciRep2022}, a mechanical oscillator in the nonlinear regime~\cite{MontenegroPRR2020}, micromechanical resonators~\cite{BrunelliPRA2011}, Bose-Einstein condensates~\cite{SabinSciRep2014}, ultracold atoms~\cite{JohnsonPRA2016}, cold Fermi gases~\cite{MitchisonPRL2020}, and even biological applications with cells~\cite{YangACSNano2011,WuAdvancedScience2022}.

In this paper, we set our analysis in the context of {\it qubit thermometers}~\cite{BrunelliPRA2011,JevticPRA2015,RazavianEPJP2019}, which have been experimentally tested in Refs.~\cite{ThamSciRep2016,MancinoPRL2017} on a quantum optics platform.
In particular, we assume that asymptotically (i.e., in the large-time limit) the quantum thermometer is in the thermal state $\rho_{\beta} = e^{-\beta H} / Z_{\beta}$, with $\beta$ denoting the inverse temperature of the bath, $H$ the Hamiltonian of the thermometer and $Z_{\beta}$ the corresponding partition function. Then, as in~\cite{MancinoPRL2017}, we consider that the thermometer weakly interacts with the thermal bath, so that the thermalization dynamics---to which the thermometer is subject---is well-described by a Markovian master equation in Gorini-Kossakowski-Sudarshan-Lindblad (GKSL) form~\cite{Lidar2019}. Because of the thermalization dynamics we study, resulting in an asymptotic thermal state regardless of the initial state, the time-evolved state of the thermometer always encodes information about the temperature that we aim to infer.

In the conditions drawn above, in order to infer the temperature $T$ of the thermal bath, one could wait for the full thermalization of the thermometer (i.e., wait for the state of the thermometer being thermal), and then reconstruct $T$ from its measure. However, the time required by the thermometer to thermalize can be very large, with the consequence that in a nanoscopic setting other sources of error probably arise. This would have the effect of disturbing the state of the thermometer, spoiling the information on $T$.

In this paper we follow a nonequilibrium approach for quantum thermometry~\cite{FeylesPRA2019} that relies on measuring the (time-dependent) state of the thermometer, while the thermalization is still active. From these measurements, the temperature of the bath is reconstructed. To guide this inference of the temperature with the best accuracy (or sensitivity) possible, we compute the quantum Fisher information, which depends on time, on the initial state of the thermometer (before it is put in contact with the thermal bath) and on the parameters of the GKSL master equation. The quantum Fisher information is a proper quantifier for also evaluating the accuracy of quantum thermometry. This is because for any unbiased estimator (in our case, the measurements of the quantum thermometer's state w.r.t.~the Hamiltonian basis), the uncertainty of the estimate (here, the reconstructed temperature) is bounded from below by the quantum Fisher information, according to the quantum Cramér-Rao bound~\cite{BraunsteinPRL1994,CorreaPRL2015,LiuJPAMT2020,YuNpjQI2022}. Such a bound is tighter the larger is the number of independent experiments that are performed to estimate $T$.

Accordingly, the aim of the paper is to look for both the time in the transient of the thermalization dynamics and the initial state of the thermometer, such that the quantum Fisher information is maximized. The result of this optimization is expected to guide the experimentalist to set the optimal conditions allowing to carry out the thermometry task with high accuracy. 
In this regard, notice that the computation of the quantum Fisher information has to be done before the thermometry experiments, requiring some a priori knowledge of the thermalization dynamics. For example, in the setting implemented experimentally in \cite{MancinoPRL2017}, which we consider in the following, one implicitly assumes that the thermalization dynamics of the quantum thermometer is well-described by a master equation in Markovian regime.

We conclude the paper by discussing a possible experimental test of our results on the quantum optics platform in Ref.~\cite{MancinoPRL2017}, and we provide some outlook for possible future works.

\section{Nonequilibrium quantum thermometer}\label{sec:model}

In this section we introduce the model of a thermometer as a $N$-level quantum system, interacting with a thermal bath in the weak-coupling regime. The following assumptions are taken: 
1) the initial state of the thermometer and bath are uncorrelated; 
2) the action of the thermometer on the bath is negligible, so that the bath remains always in a thermal state; 
3) the rotating wave approximation is valid: fast oscillating terms in the thermometer-bath dynamics, when compared to the thermometer time scale, are neglected.
As a result, after tracing out the environment degrees of freedom, the dynamics of the quantum thermometer is governed by a Lindblad master equation~\cite{Lidar2019,Palao2001}:
\begin{equation}
\label{eq:me_thermal_dyn}
  \dot{\rho}(t) = -i\left[ H, \rho(t) \right] + \sum_{i,j=1;\,\,i\neq j}^{N} \left( \mathcal{L}_{ij}\rho(t)\mathcal{L}_{ij}^{\dag} - \frac{1}{2}\{\mathcal{L}_{ij}^{\dag} \mathcal{L}_{ij},\rho(t) \}\right),
\end{equation}
where $\hbar$ is set to $1$, and the Hamiltonian is
\begin{equation} 
\label{eq:H0}
    H = \sum_{j=1}^N \epsilon_j |\epsilon_j\rangle\!\langle\epsilon_j|,
\end{equation}
with eigenvalues $\{\epsilon_j\}_{j=1}^N$ arranged in order of increasing energy, and eigenvectors $\{\ket{\epsilon_j}\}_{j=1}^N$. In Eq.~(\ref{eq:me_thermal_dyn}), the thermalization dynamics induced on the thermometer by the interaction with the thermal bath is described via the jump operators $\mathcal{L}_{ij} \equiv \sqrt{\Gamma_{ij}} |\epsilon_i\rangle\!\langle \epsilon_j|$~\cite{Scovil1959,Palao2001,Gherardini2021}.
The transition rates $\Gamma_{ij}$ from state $j$ to state $i$ are given by
\begin{align}\label{eq:eta_ij}
    \Gamma_{ij} = \begin{cases}
    \gamma (n_{ij}+1) &\quad \text{for} \quad i<j \\
    0 &\quad \text{for} \quad i=j \\
    \gamma\ n_{ji} &\quad \text{for} \quad i>j
    \end{cases} 
\end{align}
with $\gamma > 0$ having dimension of [time]$^{-1}$, and $n_{ij}$ denoting the thermal ratios
\begin{align}
\label{eq:n_ij}
    n_{ij} \equiv n_{ij}(\beta) = \frac{1}{e^{\beta \omega_{ij}} - 1} \,,
\end{align}
where $\omega_{ij} \equiv \epsilon_j - \epsilon_i$ and $\omega_{ij} > 0$ for $i<j$.
Since Eq.~(\ref{eq:me_thermal_dyn}) is a differential equation depending linearly on $\rho$, we can rewrite it as a system of linear differential equations for the terms $\rho_{ij} \equiv \langle \epsilon_i| \rho | \epsilon_j \rangle$ representing the projections of $\rho$ on the energy eigenbasis:
\begin{align}
    \dot{\rho}_{ii} &= \sum_{k=1}^N \Big( \Gamma_{ik} \rho_{kk} - \Gamma_{ki} \rho_{ii} \Big) \,, 
    \label{eq:rho_dot_ii_N}\\
    \dot{\rho}_{ij} &= \left[ -\frac{1}{2} \sum_{k=1}^N \big( \Gamma_{ki} + \Gamma_{kj} \big) + \text{i}\, \omega_{ij} \right] \rho_{ij} \quad \text{for} \quad i\neq j \,.
    \label{eq:rho_dot_ij_N}
\end{align}
The time-evolution of the diagonal population terms is decoupled from the evolution of the off-diagonal coherence ones. Hence, we will address the dynamics of each of them separately. We stress that this is due to the specific themalization dynamics we have considered, i.e., the master equation \eqref{eq:me_thermal_dyn}  with the jump operators $\mathcal{L}_{ij} \equiv \sqrt{\Gamma_{ij}} |\epsilon_i\rangle\!\langle \epsilon_j|$.\\

\noindent \textbf{Diagonal population terms}\\
\noindent
Let us denote the vector with the population terms as $\mathbf p_\mathrm{d} \equiv \begin{bmatrix} \rho_{ii} \end{bmatrix}_i$. Eq.~(\ref{eq:rho_dot_ii_N}) forms a linear differential system of $N$ equations~\footnote{
Out of these $N$ differential equations, only $N-1$ are linearly independent, since $A_\beta$ has $N-1$ non-null eigenvalues, as demonstrated in Appendix~\ref{app:Aeig}.}:
\begin{align}
\label{eq:A}
    \dot{\mathbf p}_\mathrm{d} = A_\beta\, \mathbf p_\mathrm{d}, 
\end{align}
where $A_\beta \equiv [a_{ij}]_{ij}$ is the $N\times N$ transition matrix whose entries are given by
\begin{align}
    a_{ij} &= \Gamma_{ij}\,,\label{eq:a_ij}\\
    a_{ii} &= -\sum_{k=1}^N \Gamma_{ki}\,.\label{eq:a_ii}
\end{align}
Notice that Eq.~\eqref{eq:a_ii} is a direct consequence of probability conservation in terms of the normalization of $\rho$ for any time: $\sum_{i}\rho_{ii}=1$.
Moreover, the dependence of the matrix $A_\beta$ on $\beta$ is made explicit through Eqs.~(\ref{eq:eta_ij})-(\ref{eq:n_ij}). Hence, solving the differential system (\ref{eq:A}), one gets:
\begin{align}
   \mathbf p_\mathrm{d}(t,\beta) \equiv  \mathrm{e}^{A_\beta t}  \mathbf p_\mathrm{d}(0). \label{eq:rho_diag}
\end{align}
As a consistency check, in Appendix~\ref{app:A_th}, we show that the matrix $A_\beta$ has a single null eigenvalue, while the remaining $N-1$ eigenvalues are strictly negative. In fact, by direct substitution, the vector $\boldsymbol{\pi}(\beta) \equiv \begin{bmatrix} e^{-\beta \epsilon_{k}}/ Z \end{bmatrix}_k$ containing the thermal populations $\pi_k \equiv \pi_k(\beta) = e^{-\beta\epsilon_k}/Z_{\beta}$ with $Z_{\beta} \equiv \sum_{k}e^{-\beta\epsilon_k}$, is the eigenvector of $A_\beta$ associated with the null eigenvalue: $A_\beta\,\boldsymbol{\pi}(\beta) =0$. This implies that, as $t \rightarrow \infty$, all the other $N-1$ eigenvectors of $A_\beta$ go to zero exponentially fast, so that the diagonal terms converge asymptotically to the thermal distribution.\\

\noindent \textbf{Off-diagonal coherence terms}\\
\noindent
In Eq.~(\ref{eq:rho_dot_ij_N}), each of the pairs composed by $\dot{\rho}_{ij}$ and $\dot{\rho}_{ji}$ for $i\neq j$   consists of two complex conjugate, and thus dependent differential equations. So, the system of equations [Eq.~(\ref{eq:rho_dot_ij_N})] is comprised of the $N(N-1)/2$ independent equations given by
$\dot \rho_{ij} = (-c_{ij} + \text{i}\,\omega_{ij}) \rho_{ij}$, with $c_{ij} \equiv \frac{1}{2} \big( \sum_{k=1}^N \Gamma_{ki} + \Gamma_{kj} \big) > 0$. Hence, given the initial quantum coherence term $\rho_{ij}(0)$, the time-evolution of $\rho_{ij}$ is:
\begin{align}
    \rho_{ij}(t) &= e^{-c_{ij}t} e^{\text{i}\omega_{ij}t} \rho_{ij}(0) \,.\label{eq:rho_ij}
\end{align}
Accordingly, the modulus of the off-diagonal terms, namely $|\rho_{ij}(t)|=e^{-c_{ij}t}|\rho_{ij}(0)|$, vanishes exponentially fast with decay rate $c_{ij}$. The dependence of $c_{ij}$ on $\beta$ is evidenced through Eqs.~(\ref{eq:eta_ij})-(\ref{eq:n_ij}). This implies that $\rho(t)$ asymptotically converges to a diagonal state, which is thermal in this case-study.

\subsection{Qubit thermometer}

We now focus the analysis of the thermometer dynamics on the case where the thermometer is a 2-level quantum system (thus, $N=2$). As shown in Appendix~\ref{app:A_th}, the transition matrix $A_{\beta}$ can be written in terms of the thermal distribution of the fixed-point of the thermalization map:
\begin{equation}
    A_{\beta} = \frac{\gamma}{\pi_1-\pi_2}\begin{bmatrix}
        -\pi_2 & \pi_1 \\[2.0 ex]
        \pi_2 & -\pi_1
    \end{bmatrix},
\end{equation}
where, we recall, $\pi_k \equiv e^{-\beta\epsilon_k}/Z_{\beta}$ with $\sum_k \pi_k = 1$ and $\epsilon_i \leq \epsilon_j$ for $i < j$. The eigensystem of $A_{\beta}$ is given by the pairs eigenvalue-eigenvector $\left\{ \lambda_0 = 0; \, \mathbf v_0 = \boldsymbol{\pi} = [\pi_k]_k\right\}$ and $\left\{ \lambda = \gamma (\pi_2-\pi_1)^{-1} < 0; \,\mathbf v_1 = [-1; 1]\right\}$, where the dependence of $\pi_1$, $\pi_2$ and $\lambda$ on $\beta$ is here omitted for better readability. Notice that the notation $[\cdot;\cdot]$ stands for column vector.

We introduce the operators $S \equiv [\mathbf v_0, \mathbf v_1]$ and $\Lambda \equiv {\rm diag}(0,\lambda)$ from the spectral decomposition of $A_{\beta}$ such that $A_{\beta} = S\Lambda S^{-1}$. Thus, after exponentiation, one can determine that
\begin{align} 
    e^{A_{\beta}t} &= S e^{\Lambda t} S^{-1} = \begin{bmatrix}
        \pi_1 & -1\\
        \pi_2 & 1
    \end{bmatrix} \begin{bmatrix}
        1 & 0\\ 0 & e^{\lambda t}
    \end{bmatrix} \begin{bmatrix}
        1 & 1\\
        -\pi_2 & \pi_1
    \end{bmatrix} \nonumber\\ &= 
    \begin{bmatrix}
        1-\pi_2 \bigl(1-e^{\lambda t}\bigr) & (1-\pi_2)\bigl(1-e^{\lambda t}\bigr)\\
        \pi_2 \bigl(1-e^{\lambda t}\bigr) & e^{\lambda t}+\pi_2 \bigl(1-e^{\lambda t}\bigr)
    \end{bmatrix}.
\end{align}

Consequently, the diagonal elements $\mathbf p(t,\beta,a)$ of the qubit thermometer's state at the generic time $t$ are
\begin{align}
    \mathbf p(t,\beta,a) &= e^{A_{\beta}t}\mathbf{p}_0(a) = \begin{bmatrix}
        \pi_1-e^{\lambda t}\bigl(\pi_1-(1-a)\bigr)\\
        \pi_2-e^{\lambda t}\bigl(\pi_2-a\bigr)
    \end{bmatrix} \nonumber\\ 
    &= \left( 1- e^{\lambda t} \right)\boldsymbol{\pi} + e^{\lambda t}\mathbf{p}_0(a) \,, \label{eq:p_t_beta}   
\end{align}
where $\mathbf{p}_0(a) \equiv [1-a; a]$ is the vector collecting the diagonal elements of the initial (at time $t=0$) state of the qubit thermometer.

On the other hand, the quantum coherence term $\rho_{12}(t,\beta)$ reads as
\begin{equation} \label{eq:rho12_t}
     \rho_{12}(t,\beta) = e^{\lambda t/2} e^{\text{i}\,\omega_{12}t} \rho_{12}(0),
\end{equation}
where $\rho_{12}(0)$ is the value at $t=0$.

We also show the analytical expression of the derivative of the qubit thermometer's state w.r.t.~$\beta$:
\begin{align}
    \partial_\beta \left( \mathbf{p}(t,\beta,a) \right) &= (1-\pi_2)\,\omega_{12}\,\delta(t,\beta,a)\, \mathbf{v}_1, \label{eq:partial_p}\\
    \partial_\beta \left( \rho_{12}(t,\beta) \right) &=  \alpha(t,\beta) \, \rho_{12}(t,\beta), \label{eq:partial_rho12}
\end{align}
with 
\begin{align}
    \alpha(t,\beta) &\equiv -(1-\pi_2)\pi_2\,\omega_{12} \lambda^2 t \label{eq:def_alpha}\\
    \delta(t,\beta,a) &\equiv  1-e^{\lambda t} + 2t\lambda^2 e^{\lambda t}(\pi_2-a)\,.\label{eq:def_delta}
\end{align}
The derivation of Eqs.~\eqref{eq:partial_p} and \eqref{eq:partial_rho12} is given in Appendix \ref{app:partial_deriv}, and will be used to determine the quantum Fisher information. For the sake of clarity, in Appendix \ref{appendix_sec:partial_vs_total_deriv}, we also report a discussion on the difference between using the partial and total derivative w.r.t.~$\beta$. This becomes relevant when the initial state of the thermometer, before it is put in contact with the thermal bath, is thermal at a given inverse temperature $\widetilde{\beta}$ (not necessarily different from $\beta$).

\section{Quantum Fisher information}

From this section, we are going to address the following questions: 
1) How much information about the inverse temperature $\beta$ of a thermal bath can be extracted from a quantum thermometer in the transient of the thermalization dynamics?
2) Does an initial state $\rho_0$ with quantum coherence (w.r.t.~the basis of $H$) yield more information about $\beta$ than its diagonal counterpart $\rho_\mathrm{d}$?
For clarity, $\rho_\mathrm{d}$ is a density operator with only diagonal elements that are defined over the eigenbasis spanned by $\{|\epsilon_j\rangle\}$ with $j=1,\ldots,N$. Hence, the difference between $\rho$ and $\rho_\mathrm{d}$ is a Hermitian complex matrix $\chi$ with off-diagonal elements only, containing the quantum coherence of the initial state that makes the energy levels of the thermometer Hamiltonian interfering. Moreover, we also stress that the ultimate goal of the paper is to understand, with analytical arguments, what is the precision or sensitivity (i.e., the metrological limit) in inferring $\beta$ during the transient of the thermalization dynamics to which the quantum thermometer is subject.

The information that a quantum state, described in the general case by the density operator $\rho(t,\beta)$, has at any time $t$ about the parameter $\beta$ of the thermal bath is quantified by the Quantum Fisher Information (QFI)~\cite{Helstrom1969,Holevo1982,HyllusPRA2012,LiuJPAMT2020}. The latter is formally defined as
\begin{equation}\label{eq:QFId}
    F(\rho_{\beta}(t),\beta) = 
    {\rm Tr}\Big[\rho_{\beta}(t)L_{\beta}^2(t)\Big],
\end{equation}
where $L_{\beta}(t)$ is the {\it Symmetric Logarithmic Derivative} (SLD), and we have outlined the dependence on the inverse temperature $\beta$ (when present) by means of a subscript in $\rho_\beta(t)$ and $L_{\beta}(t)$. The definition of the SLD is implicitly given by the {\it Lyapunov equation}
\begin{equation}\label{eq:SLD_d}
    \partial_{\beta}(\rho_{\beta}(t)) = \frac{1}{2}\Big(L_{\beta}(t)\rho_{\beta}(t) + \rho_{\beta}(t)L_{\beta}(t)\Big) \equiv \frac{1}{2}\Big\{ \rho_{\beta}(t), L_{\beta}(t) \Big\},
\end{equation} 
where $\partial_{\beta}$ denotes the partial derivative w.r.t.~$\beta$, and $\{\cdot,\cdot\}$ is the anti-commutator. From \eqref{eq:QFId} and using the cyclic property of the trace, we have that QFI is also given by the relation
\begin{equation}\label{eq:QFId_2}
    F(\rho_{\beta}(t),\beta) = {\rm Tr}\Big[ \partial_{\beta}\left( \rho_{\beta}(t) \right)L_{\beta}(t) \Big]\,.
\end{equation}
By substituting \eqref{eq:SLD_d} in \eqref{eq:QFId_2}, Eq.~\eqref{eq:QFId} is recovered. As given by the quantum Cramér-Rao inequality~\cite{BraunsteinPRL1994,CorreaPRL2015,LiuJPAMT2020,YuNpjQI2022}, the QFI identifies a lower bound for the uncertainty in estimating the unknown parameter (here, $\beta$), as a function of the number $M$ of independent experiments or trials performed for such an estimation. Formally, this means that, by denoting with $\Delta\beta$ the uncertainty of the $\beta$-estimate, the quantum Cramér-Rao inequality reads as
\begin{equation}
    \Delta\beta \geq \frac{ 1 }{ M F(\rho_{\beta}(t),\beta) }\,.
\end{equation}
The quantum Cramér-Rao bound provides the ultimate precision limit allowed by quantum mechanics, as long as the employed estimator is unbiased. Nevertheless, even if the unbiasedness requirement is not fulfilled by the estimator, the maximization of the QFI leads to a sub-optimal solution for reducing the estimation uncertainty. With this spirit, in this paper we will analyze the main conditions that entails the maximization of the QFI for any time of the Markovian thermalization dynamics to which the chosen quantum thermometer is subjected.

Now, we are going to compute the QFI in three distinct scenarios with an increasing level of complexity.
1) First, we will determine the QFI of a thermal state at an inverse temperature $\widetilde{\beta}$. Notice that, from now on, we will use $\widetilde{\beta}$ whenever we need to denote an inverse temperature that is not related to the one of the thermal bath we aim to infer. 2) Then, we will derive the QFI of a diagonal state $\rho_\mathrm{d}$ w.r.t.~the basis of $H$. States of this kind are produced by the thermalization dynamics in \eqref{eq:me_thermal_dyn} in the case the quantum thermometer is initialized in a diagonal state, but not necessarily thermal. 3) We will show general properties of the QFI (about $\beta$) of a generic density operator that, compared to $\rho_\mathrm{d}$, also contains quantum coherences. This calculation is needed when, at the beginning of the thermalization dynamics \eqref{eq:me_thermal_dyn}, the quantum thermometer is initialized in the generic state $\rho(0)$.  

As a remark, the analysis below about the QFI, albeit focusing on the specific parameter $\beta$ (the inverse temperature of a thermal bath) and on the thermalization dynamics, can be applied in a more general context. In fact, the analysis works regardless of the quantum dynamics returning the state on which the QFI is computed, and the properties of the QFI we determine are valid in principle in any scenario for quantum parameter estimation. 

\subsection{QFI of a thermal quantum state}

Let us consider a thermal state of the Hamiltonian $H$ in \eqref{eq:H0}, at inverse temperature $\widetilde{\beta}$, i.e.
\begin{equation}\label{eq:rho_beta_in}
    \rho_{\widetilde{\beta}} = \frac{ e^{-\widetilde{\beta}H} }{ Z_{\widetilde{\beta}} } = \sum_{j=1}^{N}\pi_{j}(\widetilde{\beta})|\epsilon_j\rangle\!\langle\epsilon_j|\,,
\end{equation}
with $Z_{\widetilde{\beta}} \equiv {\rm Tr}[e^{-\widetilde{\beta}H}] = \sum_{j=1}^{N}e^{-\widetilde{\beta}\epsilon_j}$. 
In order to determine the expression of the QFI of the thermal state $\rho_{\widetilde{\beta}}$, we have to compute both the derivative $\partial_{\widetilde{\beta}}(\rho_{\widetilde{\beta}}(t))$ and the SLD $L_{\widetilde{\beta}}$ for the case-study in analysis. As proved in Appendix~\ref{app_sec:QFI_thermal_state}, it holds that
\begin{equation}\label{eq:partial_rho_beta_main}
    \partial_{\widetilde{\beta}}(\rho_{\widetilde{\beta}}) = \sum_{j=1}^N \partial_{\widetilde{\beta}}\pi_j(\widetilde{\beta}) |\epsilon_j\rangle\!\langle \epsilon_j| = \sum_{j=1}^{N}\left(\langle H\rangle_{\rho_{\widetilde{\beta}}} - \epsilon_j\right)\pi_j(\widetilde{\beta}) |\epsilon_j\rangle\!\langle\epsilon_j|\,, 
\end{equation}
where $\langle H\rangle_{\rho_{\widetilde{\beta}}} \equiv {\rm Tr}[\rho_{\widetilde{\beta}}H] = \sum_{j=1}^{N}\epsilon_j\pi_j(\widetilde{\beta})$ is the expectation value of the Hamiltonian of the quantum thermometer w.r.t.~the thermal state $\rho_{\widetilde{\beta}}$. In this way, given the expressions of $\rho_{\widetilde{\beta}}$ and $\partial_{\widetilde{\beta}}(\rho_{\widetilde{\beta}})$, we get the SLD $L_{\widetilde{\beta}}$ that is given by the following diagonal matrix:  
\begin{equation}\label{eq:SLD_thermal_state}
    L_{\widetilde{\beta}} = \sum_{j=1}^N \left( \langle H\rangle_{\rho_{\widetilde{\beta}}} - \epsilon_j\right) |\epsilon_j\rangle\!\langle\epsilon_j|\,.
\end{equation}
The validity of Eq.~(\ref{eq:SLD_thermal_state}) can be directly verified by substituting (\ref{eq:SLD_thermal_state}) in the Lyapunov equation (\ref{eq:SLD_d}).
As a result, the QFI about the inverse temperature $\widetilde{\beta}$ of the thermal state $\rho_{\widetilde{\beta}}$ is
\begin{equation}\label{eq:QFI}
    F(\rho_{\widetilde{\beta}},\widetilde{\beta}) = \sum_{j=1}^N \left(\langle H\rangle_{\rho_{\widetilde{\beta}}} - \epsilon_j\right)^2 \pi_j(\widetilde{\beta}) \equiv {\rm Var}(H)_{\rho_{\widetilde{\beta}}} \,,
\end{equation}
where ${\rm Var}(H)_{\rho_{\widetilde{\beta}}}$ is the (thermal) variance of $H$ computed w.r.t.~the thermal state $\rho_{\widetilde{\beta}}$.

For qubits, the thermal expectation value $\langle H\rangle_{\rho_{\widetilde{\beta}}}$ and variance ${\rm Var}(H)_{\rho_{\widetilde{\beta}}}$ [here, equal to the QFI $F(\rho_{\widetilde{\beta}},\widetilde{\beta})$] are equal to
\begin{align}
    \langle H\rangle_{\rho_{\widetilde{\beta}}} &= \epsilon_1 \pi_1(\widetilde{\beta}) + \epsilon_2 \pi_2(\widetilde{\beta}) = \epsilon_1 + \omega_{12}\pi_2(\widetilde{\beta}) = \epsilon_2 - \omega_{12}(1-\pi_2(\widetilde{\beta})) \\
    {\rm Var}(H)_{\rho_{\widetilde{\beta}}} &= \omega_{12}^2 \pi_2(\widetilde{\beta}) \left( 1-\pi_2(\widetilde{\beta}) \right),\label{eq:QFI_n2}
\end{align}
where $\omega_{12}=\epsilon_2-\epsilon_1$ is the spectral gap of the qubit Hamiltonian.

Given that $F(\rho_{\widetilde{\beta}},\widetilde{\beta})$ quantifies the information contained in $\rho_{\widetilde{\beta}}$ about the inverse temperature $\widetilde{\beta}$, it is worth asking what is the Hamiltonian $H$ that maximizes $F(\rho_{\widetilde{\beta}},\widetilde{\beta})$. This problem, for a thermal state, has been already studied in \cite{CorreaPRL2015}, where it is explicitly stated that determining the spectrum of $H$ with the largest possible variance at thermal equilibrium directly entails the maximization of the sensitivity to a temperature. It is also shown that the solution to this problem (i.e., the maximization of ${\rm Var}(H)_{\rho_{\widetilde{\beta}}}$) is provided by taking the energy spectrum of an effective two-level quantum system with energies $E_{-}$ and $E_{+}$ associated, respectively, to $N_{-}$ and $N_{+} \equiv N - N_{-}$ times degenerate ground and excited states. In this way, ${\rm Var}(H)_{\rho_{\widetilde{\beta}}}$ is maximized in the case the degeneracy of the excited state is the largest possible, which is obtained by setting $N_{-}=1$. These considerations, of course, hold independently on the estimation algorithm one employs to estimate $\widetilde{\beta}$.

\subsection{QFI with an initial diagonal state}\label{sec_QFI_diagonal_state}

Let us now provide the formal expression of the QFI about an inverse temperature $\beta$ of a density operator $\rho_{\beta,{\rm d}}$ with only diagonal elements. If the initial quantum state of the thermometer is mixed w.r.t.~the eigenbasis of $H$, then no quantum coherence in such a basis arises, with the result that the state of the thermometer remains mixed for any time $t$ as given by Eq.~(\ref{eq:rho_diag}). Hence, the expression of QFI discussed in this subsection can be applied to the state of the quantum thermometer at any time $t$ of its dynamics, provided the thermometer is initialized in a mixed quantum state. Let us also remark that, while the thermometer may be initialized in a quantum state which bears no dependency on $\beta$, such a dependency is expected to arise as a consequence of the thermalization dynamics.

Using Eq.~\eqref{eq:QFId}, the QFI of the generic mixed quantum state $\rho_{\beta,{\rm d}}$ is 
\begin{equation}\label{eq:Fisher_diagonal_state}
F(\rho_{\beta,{\rm d}},\beta) = {\rm Tr}\left[ \partial_\beta(\rho_{\beta,{\rm d}}) L_{\beta,{\rm d}} \right] = {\rm Tr}\left[ \rho_{\beta,{\rm d}}L_{\beta,{\rm d}}^2 \right],    
\end{equation}
where $L_{\beta,{\rm d}}$ denotes the SLD for the case-study of deriving the QFI of a mixed quantum state. The SLD $L_\mathrm{d}$ is implicitly defined by the Lyapunov equation $\partial_\beta(\rho_{\beta,{\rm d}})=\left\{ \rho_{\beta,{\rm d}}, L_{\beta,{\rm d}} \right\}/2$. At this point, it is worth observing that, being $\rho_{\beta,{\rm d}}$ provided by a matrix with only diagonal elements, also the corresponding derivative $\partial_\beta(\rho_{\beta,{\rm d}})$ w.r.t.~$\beta$ is a diagonal matrix, as well as the SLD $L_{\beta,{\rm d}}$.
Therefore, 
\begin{equation}\label{eq:product_Ld_rhod}
    L_{\beta,{\rm d}}\,\rho_{\beta,{\rm d}} = {\rm diag}\Big( \big\{ L_{\beta,{\rm d}}(k)\rho_{\beta,{\rm d}}(k) \big\}_{k=1}^{N} \Big) = \partial_\beta(\rho_{\beta,{\rm d}})\,,
\end{equation}
where ${\rm diag}(\cdot)$ denotes a diagonal matrix whose diagonal is the vector $(\cdot)$, and $L_{\beta,{\rm d}}(k)$, $\rho_{\beta,{\rm d}}(k)$ are the $k$-th elements on the diagonal of $L_{\beta,{\rm d}}$ and $\rho_{\beta,{\rm d}}$ respectively. From Eq.~\eqref{eq:product_Ld_rhod}, the following relation follows directly, providing the formal expression for the diagonal elements of $L_{\beta,{\rm d}}$, i.e., 
\begin{equation}
    L_{\beta,{\rm d}}(k) = \frac{ \partial_\beta(\rho_{\beta,{\rm d}}(k)) }{ \rho_{\beta,{\rm d}}(k) }\,,\quad k=1,\dots,N \,.
\end{equation}
As a result,
\begin{equation}
    F(\rho_{\beta,{\rm d}},\beta) = \sum_{k} \frac{ \Big( \partial_\beta\left(\rho_{\beta,{\rm d}}(k)\right) \Big)^2 }{ \rho_{\beta,{\rm d}}(k) }\,.
\end{equation}

Before moving forward, we stress that the analytic solution of the optimization problem returning the initial state that maximizes the QFI in the case of diagonal and generic density operators is postponed to the next section, Sec.~\ref{sec:qubit_thermometer}, which deals with qubit thermometers.

\subsection{QFI with a generic initial state}

In this subsection we are going to address the following question, for any quantum thermometer undergoing the dynamics in Sec.~\ref{sec:model}: ``Does a density operator $\rho_{\beta}$ with quantum coherence, w.r.t.~the eigenbasis of $H$, yield more information about the inverse temperature $\beta$ of a thermal bath than its classical counter-part $\rho_{\beta,\mathrm{d}}$?

For this purpose, let us write the generic density operator $\rho_{\beta}$ as $\rho_{\beta} = \rho_{\beta,\mathrm{d}} + \rho_{\beta,\mathrm{coh}}$, where $\rho_{\beta,\mathrm{d}}$ is the diagonal density operator introduced in Sec.~\ref{sec_QFI_diagonal_state}, while $\rho_{\beta,\mathrm{coh}}$ is a null-diagonal operator (namely an hollow matrix) with off-diagonal elements, representing quantum coherence. For the sake of a simpler notation, in this section, we have dropped the dependency on both $t$ and $\beta$; in the remainder of the main text we will use them again whenever needed. Moreover, we recall that the time-evolution of the diagonal and off-diagonal elements of the density operator $\rho(t)$, solution of the thermalization dynamics in Sec.~\ref{sec:model}, are decoupled from each other, as given respectively by Eqs.~(\ref{eq:rho_dot_ii_N})-(\ref{eq:rho_dot_ij_N}).

Now, we are in the position to write the QFI $F(\rho,\beta)$ of the state of the quantum thermometer (at a given time $t$ that we do not specify), as composed by the QFI $F(\rho_\mathrm{d},\beta)$ associated to the diagonal elements of $\rho(t)$, plus an additional non-negative term whose expression we are going to provide. We will make abundant use of the following Proposition that can be easily proved by direct substitution: 
\begin{prp}\label{prp}  
Let $C$ be an $n\times n$ hollow matrix (i.e., its diagonal elements are equal to zero).    
Moreover, let $D$ be an $n\times n$ diagonal matrix. Then, both $CD$ and $DC$ are hollow matrices.
\end{prp}
We also introduce $L$ and $L_\mathrm{coh}$ as the SLD of $\rho$ and $\rho_\mathrm{coh}$ respectively, both defined w.r.t.~the inverse temperature $\beta$, so that
$2\partial_\beta(\rho) = \{ \rho, L\}$ and
\begin{align}\label{eq:SLD_coh} 
2\partial_\beta(\rho_\mathrm{coh}) = \{\rho_\mathrm{coh}, L_\mathrm{coh}\}\,.  
\end{align}
Moreover, we define $\widetilde{L} \equiv L - L_{\rm d}$, where $L_{\rm d}$ is the SLD for the diagonal density operator $\rho_{\rm d}$ that is implicitly defined via the Lyapunov equation (\ref{eq:SLD_d}) in Sec.~\ref{sec_QFI_diagonal_state}. It is worth observing that $\widetilde{L} \neq L_\mathrm{coh}$, since the Lyapunov equation represents a non-linear transformation for a density operator $\rho$. Such a feature becomes evident by expanding $2\left( \partial_{\beta}(\rho_{\rm d}) + \partial_{\beta}(\rho_{\rm coh}) \right) = 2\partial_{\beta}(\rho) = \{ \rho, L \} = \{ ( \rho_{\rm d} + \rho_{\rm coh} ), (  L_{\rm d}+\widetilde{L}  )\}$ that leads to the relation
\begin{equation}
    \{\rho_\mathrm{coh}, L_\mathrm{coh}\} = 
    \{ \rho_{\rm coh}, \widetilde{L} \} + \left( \{\rho_\mathrm{coh}, L_\mathrm{d}\} + \{ \rho_{\rm d}, \widetilde{L} \} \right), 
\end{equation}
which is evidently different from $\{ \rho_{\rm coh}, \widetilde{L} \}$. The derivative $\partial_\beta(\rho_\mathrm{coh})$ is a hollow operator by definition, and $\{ \rho_\mathrm{coh}, L_\mathrm{d} \}$ is hollow due to Proposition \ref{prp}. Hence, also $\{ \rho, \widetilde{L} \} = \{ \rho_{\rm coh}, \widetilde{L} \} + \{ \rho_{\rm d}, \widetilde{L} \}$ is a hollow operator.

We focused on the operator $\widetilde{L}$ given its importance for the computation of the QFI $F(\rho,\beta) = {\rm Tr}\left[ \partial_\beta(\rho) L \right] = {\rm Tr}\left[ \rho L^2 \right]$. In fact, as provided in the proof at the end of this section, we find that the QFI $F(\rho,\beta)$ can be decomposed as the corresponding ``classical'' Fisher information yielded by the diagonal density operator $\rho_{\rm d}$, plus the extra non-negative term ${\rm tr}\big(\rho\widetilde{L}^2\big)$: 
\begin{equation}\label{eq:QFI_coh}
    F(\rho, \beta) = {\rm Tr}\left[ \partial_\beta(\rho) L \right] =  F(\rho_\mathrm{d},\beta) + {\rm Tr}\left[ \rho \widetilde{L}^2 \right],  
\end{equation}
where $F(\rho_\mathrm{d},\beta) = {\rm Tr}\left[ \partial_\beta(\rho_\mathrm{d}) L_\mathrm{d} \right] = {\rm Tr}\left[ \rho_\mathrm{d} L_\mathrm{d}^2 \right]$ [see Eq.~\eqref{eq:Fisher_diagonal_state}]. Once again, it is worth pointing out that ${\rm Tr}[ \rho \widetilde{L}^2 ] \geq 0$ for any time $t$ due to the positive semi-definiteness of $\rho$ and $\widetilde{L}^2$; indeed, the eigenvalues of $\rho$ and $\widetilde{L}^2$ are non-negative. Therefore, in conclusion, the QFI acquired by a generic quantum state $\rho$ about the inverse temperature $\beta$ of a thermal bath is always {\it greater or equal} than the information acquired by a diagonal density operator $\rho_{\rm d}$ whose diagonal elements are the same as those of $\rho$. Interestingly, this analysis is not specific to the thermalization dynamics in Sec.~\ref{sec:model}, but holds for a generic open quantum map.

In the next section, we will show the analytical expression of $\widetilde{L}$ for qubit thermometers~\cite{BrunelliPRA2011,JevticPRA2015,ThamSciRep2016,MancinoPRL2017,RazavianEPJP2019}.\\ \\
\noindent
{\bf Proof of Eq.~\eqref{eq:QFI_coh}:} In order to determine the expression of the QFI $F(\rho, \beta) = {\rm Tr}\left[ \partial_\beta(\rho) L \right]$, let us evaluate the terms composing $\partial_\beta(\rho)L$, recalling that $\rho = \rho_{\rm d} + \rho_{\rm coh}$ and $L = L_{\rm d} + \widetilde{L}$:
\begin{eqnarray*}
\label{eq:drhoL}
    2 \partial_\beta(\rho)L  &=& \{\rho, L\} L = 
    \{ (\rho_\mathrm{d} + \rho_\mathrm{coh}) , L_\mathrm{d} + \widetilde{L} \} L
    \nonumber
    \\
    &=& \Big( \{ (\rho_\mathrm{d} + \rho_\mathrm{coh}) , L_\mathrm{d} \} + 
    \{ (\rho_\mathrm{d} + \rho_\mathrm{coh}) , \widetilde{L} \} \Big) L
    \nonumber
    \\
    &=& \{ \rho_\mathrm{d}  , L_\mathrm{d} \} L + 
    \{ \rho_\mathrm{coh} , L_\mathrm{d} \} L + 
    \{ \rho_\mathrm{d}  , \widetilde{L} \} L +
    \{ \rho_\mathrm{coh} , \widetilde{L} \} L 
    \nonumber
    \\
    &=& \{ \rho_\mathrm{d}  , L_\mathrm{d} \} L + 
    \{ \rho_\mathrm{coh} , L_\mathrm{d} \} L + 
    \{ \rho  , \widetilde{L} \} L
    \nonumber
     \\
     &=& \{\rho_\mathrm{d}, L_\mathrm{d}\}L_\mathrm{d} + \underbrace{\{\rho_\mathrm{d}, L_\mathrm{d}\}\widetilde{L}}_{*} + \underbrace{\{\rho_\mathrm{coh}, L_\mathrm{d}\}L_\mathrm{d}}_{\mathrm{hollow\, via\, \ref{prp}}} + \underbrace{\{\rho_\mathrm{coh}, L_\mathrm{d}\}\widetilde{L}}_{*} + \underbrace{\{\rho, \widetilde{L}\}L_\mathrm{d}}_{\mathrm{hollow\, via\, \ref{prp}}} + \{\rho, \widetilde{L}\}\widetilde{L} \,.   
\end{eqnarray*}
As also pointed out in the formula, the terms $\{\rho_\mathrm{coh}, L_\mathrm{d}\}L_\mathrm{d}$ and $\{\rho, \widetilde{L}\}L_\mathrm{d}$ are hollow matrices according to Proposition \ref{prp}. Thus, their trace is identically equal to zero. Moreover, by summing the terms identified with $*$, one gets $\{\rho_\mathrm{d}, L_\mathrm{d}\}\widetilde{L} + \{\rho_\mathrm{coh}, L_\mathrm{d}\}\widetilde{L} = \{\rho, L_\mathrm{d}\}\widetilde{L}$, whereby ${\rm Tr}[ \{ \rho, L_\mathrm{d}\}\widetilde{L} ] = {\rm Tr}[ \{\rho, \widetilde{L}\}L_\mathrm{d} ]=0$ due to using the cyclic property of the trace and again Proposition \ref{prp}. As a result, 
\begin{eqnarray*}
    F(\rho, \beta) &=& {\rm Tr}\left[ \partial_\beta(\rho)L \right] = \frac{1}{2}{\rm Tr}\left[ \{\rho_\mathrm{d}, L_\mathrm{d}\}L_\mathrm{d} \right] + \frac{1}{2}{\rm Tr}\left[ \{\rho, \widetilde{L}\}\widetilde{L} \right]=\\
    &=& {\rm Tr}\left[ \rho_\mathrm{d}L_\mathrm{d}^2 \right] + {\rm Tr}[ \rho\widetilde{L}^2 ] \,.
\end{eqnarray*}
  
\section{Metrological limits of qubit thermometers: Analytical derivation}
\label{sec:qubit_thermometer}

\subsection{Derivation of the QFI}

Let us consider a generic density operator $\rho(0)$ for a qubit, parameterized by $(a,r,\phi)\in [0,1]\times [0,1]\times [0,2\pi[$:
\begin{equation}\label{eq:qubit_state_parametrization}
    \rho(0) = \begin{bmatrix}
        1-a && \sqrt{(1-a)a}\, r\, e^{\text{i}\phi}\\
        \sqrt{(1-a)a}\, r\, e^{-\text{i}\phi} && a
    \end{bmatrix}.
\end{equation}
Thus, a qubit thermometer with initial state $\rho(0)$, undergoing the thermalization dynamics from Eq.~(\ref{eq:me_thermal_dyn}) with population terms as in Eq.~(\ref{eq:p_t_beta}) and with coherence decay rate $c_{12} = c_{21} = \frac12\bigl(\Gamma_{12} + \Gamma_{21}\bigr) = -\frac12\lambda = -\frac{\gamma}{2}(\pi_2-\pi_1)^{-1}$, evolves as
\begin{align}
    \rho_{\beta}(t) &= \begin{bmatrix}
        1-\pi_2+e^{\lambda t}(\pi_2-a) && e^{\frac12\lambda t}e^{\text{i}\omega_{12}t}\rho_{12}(0)\\
        e^{\frac12\lambda t}e^{-\text{i}\omega_{12}t}\rho_{12}^{*}(0) && \pi_2-e^{\lambda t}(\pi_2-a)
    \end{bmatrix} = \nonumber\\ &\equiv \begin{bmatrix}
        1-\rho_{22}(t,\beta) && \rho_{12}(t,\beta)\\
        \rho_{12}^{*}(t,\beta) && \rho_{22}(t,\beta)
    \end{bmatrix}.
\end{align}

To study the QFI of the thermometer's state $\rho_{\beta}(t)$ about $\beta$, we need to compute the symmetric logarithmic derivative $L_{\beta}(t)$ of $\rho_{\beta}(t)$, as in Eq.~(\ref{eq:SLD_d}). To do so, we compute $\partial_\beta\left(\rho_{\beta}(t)\right)$ using Eqs.~\eqref{eq:partial_p} and \eqref{eq:partial_rho12}:
\begin{align}
    \partial_\beta \left( \rho_{\beta}(t) \right) &= \begin{bmatrix}
        -(1-\pi_2)\pi_2\, \omega_{12}\, \delta(t,\beta,a) && \alpha\rho_{12}(t,\beta) \\
        \alpha \rho_{12}^{*}(t,\beta) && (1-\pi_2)\pi_2\, \omega_{12}\, \delta(t,\beta,a)
    \end{bmatrix} = \nonumber\\ &\equiv \begin{bmatrix}
        -\partial_\beta\left(\rho_{22}(t,\beta)\right) && \alpha\rho_{12}(t,\beta)\\
        \alpha \rho_{12}^{*}(t,\beta) && \partial_\beta\left(\rho_{22}(t,\beta)\right)
    \end{bmatrix},
\end{align}
where $\delta(t,\beta,a)$ is defined by Eq.~\eqref{eq:def_delta}. 
Thus, expanding Eq.~(\ref{eq:SLD_d}) and writing $L_{\beta}(t) \equiv \begin{bmatrix}
    \ell_{11}(t,\beta) && \ell_{12}(t,\beta) \\ \ell_{12}^{*}(t,\beta) && \ell_{22}(t,\beta)
\end{bmatrix}$, we end-up with the following three equations for three unknowns:
\begin{align}
    -\partial_\beta\left(\rho_{22}(t,\beta)\right) &= \big( 1-\rho_{22}(t,\beta)\big)\ell_{11}(t,\beta) + \textrm{Re}\left\{ \rho_{12}^{*}(t,\beta)\, \ell_{12}(t,\beta) \right\} \label{eq:equ_ell_1} \\
    \partial_\beta\left( \rho_{22}(t,\beta) \right) &= \rho_{22}(t,\beta)\, \ell_{22}(t,\beta) + \textrm{Re}\left\{ \rho_{12}^{*}(t,\beta)\, \ell_{12}(t,\beta) \right\} \\
    2 \alpha \, \rho_{12}(t,\beta) &= \big( \ell_{11}(t,\beta) + \ell_{22}(t,\beta) \big) \rho_{12}(t,\beta) + \ell_{12}(t,\beta). \label{eq:equ_ell_3}
\end{align}
Solving Eqs.~\eqref{eq:equ_ell_1}-\eqref{eq:equ_ell_3} as a function of $\ell_{11}(t,\beta)$, $\ell_{22}(t,\beta)$ and $\ell_{12}(t,\beta)$ leads us to:
\begin{align}
    \ell_{11} &= \frac{ 2\partial_\beta\left( \rho_{22} \right) |\rho_{12}|^2 - 2\alpha\, \rho_{22}|\rho_{12}|^2 - \rho_{22}\, \partial_\beta\left( \rho_{22} \right) }{ (1-\rho_{22})\rho_{22} - |\rho_{12}|^2 } \label{eq:explicit_equ_ell_1} \\
    \ell_{22} &= \frac{ -2\partial_\beta\left(\rho_{22}\right)|\rho_{12}|^2 - 2\alpha (1-\rho_{22}) |\rho_{12}|^2 + (1-\rho_{22}) \partial_\beta\left( \rho_{22} \right) }{ (1-\rho_{22})\rho_{22} - |\rho_{12}|^2 } \\
    \ell_{12} &= \frac{ 2\alpha (1-\rho_{22})\rho_{22} - (1-2\rho_{22}) \partial_\beta\left( \rho_{22} \right) }{ (1-\rho_{22})\rho_{22} - |\rho_{12}|^2 }\, \rho_{12}\,. \label{eq:explicit_equ_ell_3}
\end{align}
As a result, the QFI at time $t$ of the thermometer's state $\rho_{\beta}(t)$ about $\beta$ is:
\begin{align}\label{eq:QFI_t}
    F(\rho_{\beta}(t),\beta) &= {\rm Tr}\left[ \partial_\beta\left(\rho_{\beta}(t)\right) L_{\beta}(t)\right] = \partial_\beta\left(\rho_{22}\right)\left( \ell_{22} - \ell_{11} \right) + 2 \alpha \textrm{Re}\left\{ \rho_{12}^{*} \, \ell_{12} \right\} = \nonumber\\
    &= \frac{4\alpha |\rho_{12}|^2 \Big( \alpha (1-\rho_{22})\rho_{22} + (1-2\rho_{22}) \partial_\beta \rho_{22} \Big) + \left(\partial_\beta\left(\rho_{22}\right)\right)^2 \bigl(1-4|\rho_{12}|^2\bigr) }{ (1-\rho_{22})\rho_{22} - |\rho_{12}|^2 } = \nonumber\\
    &= \frac{ \left(\partial_\beta\left(\rho_{22}\right)\right)^2 }{ (1-\rho_{22})\rho_{22} - |\rho_{12}|^2 } + 4|\rho_{12}|^2 \frac{ \alpha^2 (1-\rho_{22})\rho_{22} + \alpha (1-2\rho_{22}) \partial_\beta \left( \rho_{22} \right) - \left( \partial_\beta \left(\rho_{22} \right)\right)^2 }{ (1-\rho_{22})\rho_{22} - |\rho_{12}|^2 }\,, 
\end{align}
where $\alpha$ is given by Eq.~\eqref{eq:def_alpha}, and all the elements of $L_{\beta}(t)$ and $\rho_{\beta}(t)$ in Eqs.~\eqref{eq:explicit_equ_ell_1}-\eqref{eq:explicit_equ_ell_3} and Eq.~\eqref{eq:QFI_t} depend on $t$ and $\beta$ (albeit not explicitly written).

Now, some remarks are in order. 1) If the initial density operator $\rho(0)$ is diagonal (i.e., $r=0$), then $\rho_{12}(t,\beta)=0$ for any time $t$ and the QFI of $\rho_{\beta}(t)$ about $\beta$ is equal to the QFI of the diagonal density operators with elements $\mathbf p(t,\beta,a)$ that one obtains by initializing the quantum thermometer in ${\rm diag}(1-a,a)$. 2) The QFI $F(\rho_{\beta}(t),\beta)=0$ at $t=0$, given that $\alpha(0,\beta)=0$ and $\delta(0,\beta,a)=0$ for any $a,\beta$. Thus, as expected, measuring the state of the quantum thermometer at $t=0$ yields no information on the inverse temperature of the thermal bath, as the initial state of the thermometer is $\beta$-independent. 3) For $t\rightarrow \infty$, $\delta \rightarrow 1$ and $|\rho_{12}|$ vanishes exponentially fast; hence, the QFI is converging to a $\beta$-dependent value that is the thermal variance of the Hamiltonian $H$ [see Eq.~(\ref{eq:QFI_n2})]. 4) The diagonal and off-diagonal elements of the quantum thermometer's state always refer to the Hamiltonian $H$. Thus, the latter implicitly represents the observable that we are considering to measure via projective measurements (i.e., projections on the eigenbasis $\{ \ketbra{\epsilon_1}{\epsilon_1}, \ketbra{\epsilon_2}{\epsilon_2}\}$), to carry out thermometry in a nonequilibrium regime. It is the optimal solution, over all the possible measurement observable (even $\beta$-dependent) if the initial state of the thermometer is diagonal in $H$.

\begin{figure}[t]
    \centering
    \includegraphics[width=0.6\columnwidth]
    {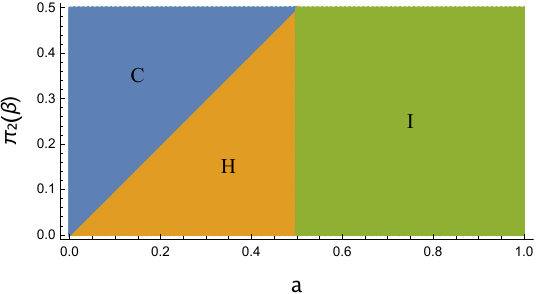}
    \caption{Pictorial representation of the parameter region identifying three distinct time-behaviors of the QFI depending on the value of both $\pi_2(\beta)\in[0,\frac{1}{2}]$ and $a\in[0,1]$.}
    \label{fig:FI2_phase_space}
\end{figure}

\subsection{The role of the diagonal elements in the initial density operator}

To analyze the time-behavior of the QFI, we start by considering initial density operators with only diagonal elements $a$ and $1-a$ (thus $r=0$), with $a\in[0,1]$. In this regard, it is worth noting that one can determine three different regions for the QFI depending on the value of the temperature of the thermal bath, and the parameter $a$ that characterizes the initial state of the qubit thermometer.

Once fixed the energies $\epsilon_1$ and $\epsilon_2$ of the thermometer, the temperature $\beta$ (taken as positive) uniquely defines the thermal probability $\pi_2(\beta)$, as well as $\pi_1(\beta)=1-\pi_2(\beta)$. For the scope of our analysis, $\pi_2(\beta)$ is considered belonging to the interval $[0,\frac{1}{2}]$. Moreover, for a qubit, any density operator with diagonal elements can be written as a thermal state. Accordingly, if $a\in [0,\pi_2[$, then the initial state of the thermometer is associated to a thermal distribution with a colder temperature than $\boldsymbol{\pi}$. This fact gives a specific behavior to the time-evolution of the QFI. In Fig.~\ref{fig:FI2_phase_space}, the region of the parameter space corresponding to $a\in [0,\pi_2[$ is denoted as `Region $\mathcal C$'. 

Then, any initial density operator with $a\in\ ]\pi_2,\frac{1}{2}]$ can be regarded as a thermal state with a hotter temperature than $\boldsymbol{\pi}$; in Fig.~\ref{fig:FI2_phase_space}, we denote such a parameter region as `Region $\mathcal H$'.

\begin{figure}[t!]
    \centering
    \begin{subfigure}[t]{0.485\textwidth}
        \centering
        \includegraphics[width=\textwidth]{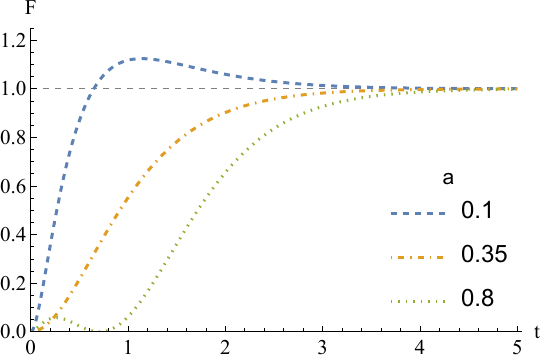}
        \caption{Time-behaviour of the QFI for initial diagonal density operators with $a\in\{0.1,0.35,0.8\}$. Each of the values of $a$ is related to a different region: respectively, the regions $\mathcal C$, $\mathcal H$, and $\mathcal I$ depicted in Fig.~\ref{fig:FI2_phase_space}.}
        \label{fig:FI_var_a}
    \end{subfigure}
    \hfill
    \begin{subfigure}[t]{0.485\textwidth}
        \centering
        \includegraphics[width=\textwidth]{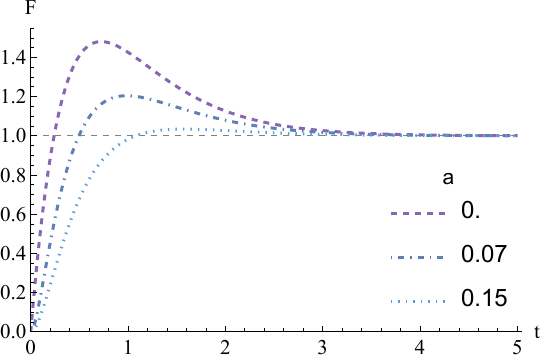}
        \caption{Time-behaviour of the QFI for initial diagonal states colder than the thermal state $\boldsymbol{\pi}$ (region $\mathcal C$ in Fig.~\ref{fig:FI2_phase_space}).}
        \label{fig:FI_cold}
    \end{subfigure}
    \vskip\baselineskip
    \begin{subfigure}[t]{0.485\textwidth}
        \centering
        \includegraphics[width=\textwidth]{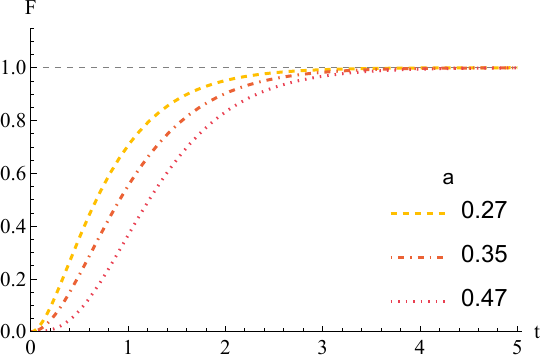}
        \caption{Time-behaviour of the QFI initial diagonal states hotter than the thermal state $\boldsymbol{\pi}$, (region $\mathcal H$ in Fig.~\ref{fig:FI2_phase_space}).}
        \label{fig:FI_hot}
    \end{subfigure}
    \hfill
    \begin{subfigure}[t]{0.485\textwidth}
        \centering
        \includegraphics[width=\textwidth]{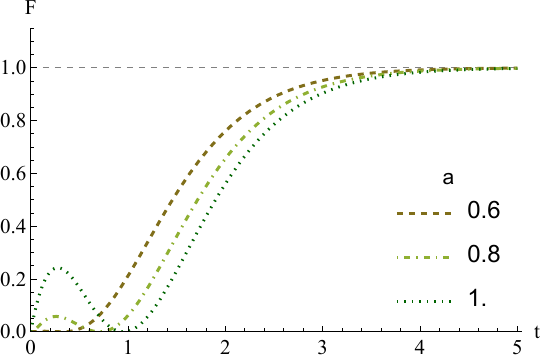}
        \caption{Time-behaviour of the QFI for inverted initial diagonal states (region $\mathcal I$ in Fig.~\ref{fig:FI2_phase_space}).}
        \label{fig:FI_inv}
    \end{subfigure}
    \caption{Time-behaviour of the QFI for initial diagonal density operators, showing the different regimes represented by the regions $\mathcal C$, $\mathcal H$, and $\mathcal I$ in Fig.~\ref{fig:FI2_phase_space}. In all figures $\pi_2=0.25$}
    \label{fig:FI_var_a}
\end{figure}

Finally, any initial diagonal state of the qubit thermometer with $a\in\ ]\frac{1}{2},1]$ can be related to a thermal state with an `effective negative temperature' that simply stands for a population-inverted state, so that the excited state is more populated than the ground state. This region is denoted as `Region $\mathcal I$' in Fig.~\ref{fig:FI2_phase_space}.

Let us now show the distinct behaviors for the time-evolution of the QFI $F$ in the three regions, which we plot altogether in Fig.~\ref{fig:FI_var_a} as a function of $a$ taking $\omega_{12}=1$, $\gamma=1$ and $\pi_2(\beta) = 0.25$ (all these quantities are expressed in dimensionless units). We detail the curves referring to `Region $\mathcal{C}$' in Fig.~(\ref{fig:FI_cold}), where we can observe that the QFI increases monotonically until it reaches a global maximum at some finite time $t^*$. After such a time, $F$ decreases monotonically to the asymptotic value corresponding to the thermal fixed point of the thermalization dynamics. In all the panels of Fig.~\ref{fig:FI_var_a}, the QFI is normalized to such an asymptotic value, which is always the same independently on the initial state of the thermometer. In Fig.~(\ref{fig:FI_cold}), the colder the initial state (i.e., the smaller the value of $a$ with respect to $\pi_2(\beta)$), the greater is the maximum value of the Fisher information that occurs at the early time $t^*$. 

On the other hand, by initializing the qubit thermometer using parameters lying in the `Region $\mathcal{H}$', $F$ increases monotonically from $0$ to the asymptotic value, as shown in Fig.~(\ref{fig:FI_hot}). The hotter the initial state (i.e., the greater the value of $a$ in the interval $]\pi_2,\frac{1}{2}]$\,), the slower is the convergence of $F$ to the asymptotic value.

We also detail in Fig.~(\ref{fig:FI_inv}) the time-evolution of the QFI considering an initial inverted state for the qubit thermometer. In such a case, $F$ increases until a local maximum, then decreases to zero, after which it monotonically increases again until the asymptotic value.

\subsection{The role of coherence in the initial density operator}

Having analyzed the evolution of the QFI $F$ in time for different initial diagonal states, we now study the role of quantum coherence in the initial state of a qubit thermometer.
Hence, differently to what done previously, we initialize the thermometer in a pure quantum state of the form as given by Eq.~\eqref{eq:qubit_state_parametrization} with $r=1$ and $\phi=0$. 

\begin{figure}[t!]
    \centering
    \begin{subfigure}[b]{0.285\textwidth}
        \centering
        \includegraphics[width=\textwidth]{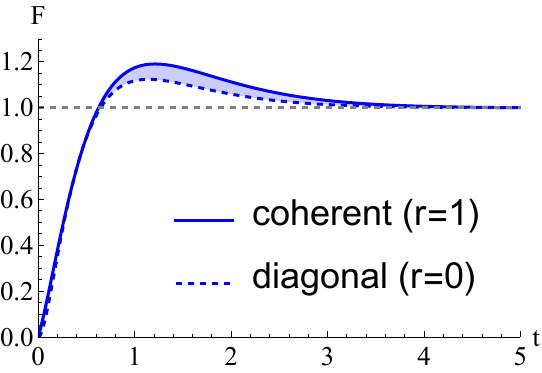}
        \caption{Cold initial state; diagonal elements: $a=0.1$.}
        \label{fig:FI_cold_coh}
    \end{subfigure}
    \hfill
    \begin{subfigure}[b]{0.285\textwidth}
        \centering
        \includegraphics[width=\textwidth]{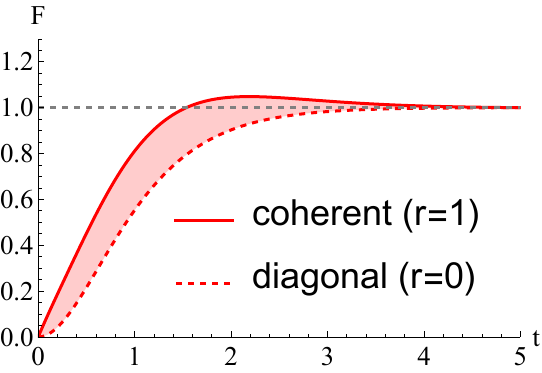}
        \caption{Hot initial state; diagonal elements: $a=0.35$.}
        \label{fig:FI_hot_coh}
    \end{subfigure}
    \hfill
    \begin{subfigure}[b]{0.285\textwidth}
        \centering
        \includegraphics[width=\textwidth]{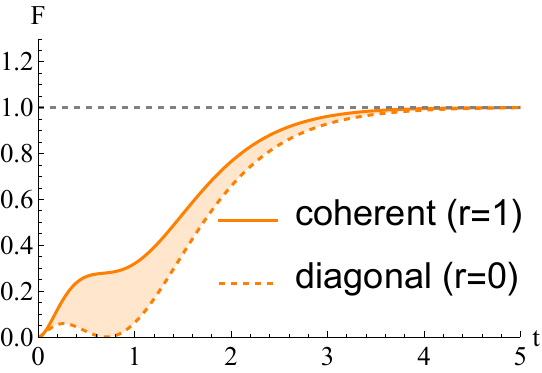}
        \caption{Inverted initial state; diagonal elements: $a=0.8$.}
        \label{fig:FI_inv_coh}
    \end{subfigure}
    \caption{QFI $F(\rho_{\beta}(t),\beta)$ as a function of time $t$, with $\pi_2(\beta)=0.25$. We compare the different behaviours given from initializing the qubit thermometer in a diagonal initial density operator, $r=0$, or in a pure initial state, $r=1$, which bears some quantum coherence. From panel to panel we vary the diagonal elements in the initialization of the thermometer.}  
    \label{fig:FI_coh_comparison}
\end{figure}

As first, notice that the quantum coherence in the initial state modifies the value of the QFI $F(\rho_{\beta}(t),\beta)$, Eq.~\eqref{eq:QFI_t}, via the term $|\rho_{12}(t,\beta)|^2$. Thus, the phase $\phi$ entering the coherence term $\sqrt{(1-a)a}\, r\, e^{\text{i}\phi}$ in $\rho(0)$ does not influence the QFI. Then, in Fig.~\ref{fig:FI_coh_comparison} we plot the time-behaviour of $F(\rho_{\beta}(t),\beta)$ for pairs of initial quantum states given by a pure state ($r=1$) and a diagonal one ($r=0$) with the same diagonal elements (thus, the same value of $a$). As a result, for the same value of $a$, setting $r=1$ (meaning that quantum coherence is present in the initial state of qubit thermometer) instead of $r=0$ brings an advantage in terms of the QFI maximization at finite times. Such an advantage due to quantum coherence decreases for $a$ small, and vanishes if $a=0$. However, even with $r=1$, the best performances in terms of magnitude of QFI for any time $t$ are obtained setting $a=0$ that refers to the ground state of $H_0$.

We conclude this section by stressing that, in the case
the qubit thermometer is initialized in a pure state ($r=1$), the three distinct behaviours of the QFI over time outlined in Fig.~\eqref{fig:FI2_phase_space} [i.e., $a\in [0,\pi_2[$ (Region $\mathcal C$), $a\in\ ]\pi_2,\frac{1}{2}]$ (Region $\mathcal H$), and $a\in\ ]\frac{1}{2},1]$ (Region $\mathcal I$)] are no longer valid in general, as the QFI in Region $\mathcal H$ can be larger than 1.

\section{Discussion}

In this paper, we have computed the QFI associated to a quantum thermometer in weak contact with a thermal bath. Maximizing the QFI allowed us to determine the optimal time, within the transient thermalization dynamics, and the initial state of the quantum thermometer such that the thermometry accuracy is enhanced. We specialized our analysis to the case of qubit thermometers, whereby analytical expressions are derived.

Now we are going to discuss the application of our results to an experimental platform. We consider the quantum optics setup in \cite{MancinoPRL2017} where the dynamics of a qubit thermometer in interaction with a thermal bath is simulated. The thermometer Hamiltonian is $H = \hbar\omega\sigma_z/2$, so that $\epsilon_2 = \hbar\omega/2$, $\epsilon_1 = -\hbar\omega/2$, and thus $\omega_{12} = \epsilon_2 - \epsilon_1 = \hbar\omega$.\\
The thermalization dynamics of the qubit in the experiments is described by a generalized amplitude damping channel, defined by the Kraus operators~\cite{MancinoPRL2017,Lidar2019}
\begin{equation}
K_0 = \sqrt{p_1}\begin{bmatrix} 1 & 0 \\ 0 & \sqrt{1-p_2} \end{bmatrix}, \quad K_1 = \sqrt{p_1}\begin{bmatrix} 0 & \sqrt{p_2} \\ 0 & 0 \end{bmatrix}, \quad K_2 = \sqrt{1-p_1}\begin{bmatrix} \sqrt{1-p_2} & 0 \\ 0 & 1 \end{bmatrix}, \quad K_3 = \sqrt{1-p_1}\begin{bmatrix} 0 & 0 \\ \sqrt{p_2} & 0 \end{bmatrix},
\end{equation}
where the probabilities $p_1$ and $p_2$ are: $p_1 = \frac{n_{12}}{2n_{12}-1}$ and $p_2 = 1 - \exp\left(-(1+n_{12})\widetilde{\tau}\right)$, with $n_{12}$ the usual thermal ratio, Eq.~(\ref{eq:n_ij}). The quantity $\widetilde{\tau}$ is the dimensionless time that is representative of the duration of the thermalization dynamics; in \cite{MancinoPRL2017}, $\widetilde{\tau}$ is taken in the interval $[0,0.3]$. It is worth noting that $p_1$ is approximately equal to $1/2$ ($p_1 \simeq 1/2$) for $n_{12} > 5$, and that $p_2 \approx (1+n_{12})\widetilde{\tau}$ for $n_{12}\widetilde{\tau} < 1$. Thus, comparing the Kraus operator $K_1$ and the jump operator $\mathcal{L}_{12}$ in Sec.~\ref{sec:model}, in first approximation we can set: $\Gamma_{12} \approx p_1 p_2 \omega_{12}$, where multiplying by $\omega_{12}$ allows to give $\Gamma_{12}$ the correct dimensionality of [time]$^{-1}$ with $\hbar=1$. Accordingly, $\Gamma_{12} \approx \frac{\widetilde{\tau}\omega_{12}}{2}(1+n_{12})$, i.e., $\gamma \approx \frac{\widetilde{\tau}\omega_{12}}{2}$. Using the generalized amplitude damping channel gives comparable results w.r.t.~the ones provided by the model we introduced in Sec.~\ref{sec:model}.\\
The qubit thermometer is initialized in the pure state $\rho(0)=\ketbra{\psi}{\psi}$ with $|\psi\rangle = \cos(\theta/2)|0\rangle + \sin(\theta/2)|1\rangle$, where $|0\rangle$ and $|1\rangle$ are the eigenstates of the Pauli matrix $\sigma_z$, and $\theta \in [0,2\pi]$. The angle $\theta$ sets the magnitude of the quantum coherence in $\rho(0)$, which is equal to $\cos(\theta/2)\sin(\theta/2)= \sin(\theta)/2$. In \cite{MancinoPRL2017}, $|0\rangle$ and $|1\rangle$ are the horizontal $|H\rangle$ and vertical $|V\rangle$ polarization states of the photons employed as thermometers, while $\theta$ is given by the birefringent angle of the spatial light modulator composing a Sagnac interferometer. According to the parametrization in (\ref{eq:qubit_state_parametrization}), the initial state of the qubit thermometer realized in \cite{MancinoPRL2017} is obtained by setting $a = \sin^2(\theta/2)$ [i.e., $\theta = 2\arcsin(\sqrt{a})$], $r = 1$ and $\phi = 0$.\\
After the initialization, the thermometer is put in contact with the thermal bath for a time $\tau$, which varied in different experiments. This means that for each experiment one chooses the time $\tau$, then lets the thermometer interact with the thermal bath for the duration $\tau$, and finally measures the state of the thermometer (via quantum state tomography), with the goal to determine the temperature of the bath.\\
In \cite{MancinoPRL2017}, a thermometry task is carried out by discriminating between two different values of $n_{12} = 1/( e^{\beta\omega_{12}} - 1)$ (dimensionless number): $n_{12}^{(c)}=5.5$ and $n_{12}^{(h)}=9.5$, corresponding respectively to the effective temperatures of a cold and hot thermal bath. Hence, the inverse temperature $\beta$ as a function of $n_{12}$ is: $\beta = \frac{1}{\omega_{12}}\ln\left( \frac{1+n_{12}}{n_{12}} \right)$, so that $\beta_h \approx 0.020$ and $\beta_c \approx 0.033$ by choosing $\omega_{12}=5$.

Referring to the experimental setting in \cite{MancinoPRL2017}, we can determine both the optimal value of $\theta$ in the initial state of the thermometer and the optimal time $t^{*}$ at which performing the thermometry, such that the analytical expression of the QFI we have computed in Sec.~\ref{sec:qubit_thermometer} is maximized. This is useful since the maximization of the QFI leads to enhance the accuracy in estimating the value of $\beta$. Thus, let us set $\omega_{12}=5$ (i.e., $\epsilon_1 = -\omega_{12}/2 = -2.5$, $\epsilon_2 = \omega_{12}/2 = 2.5$, $\beta_h \approx 0.020$ and $\beta_c \approx 0.033$), and $\widetilde{\tau}=0.05$ so that $\gamma \approx 0.125$. With this choice of the parameter values, we have that $\lambda_j = \gamma\left( \pi_2(\beta_j) - \pi_1(\beta_j) \right)^{-1}$, with $\pi_k(\beta_j) = e^{-\beta_{j}\epsilon_k}/Z_{\beta_j}$, entering in the analytical expressions of $\alpha(t,\beta_j)$ and $\delta(t,\beta_j,a(\theta))$ of Eqs.~\eqref{eq:def_alpha}-\eqref{eq:def_delta} with $j=c,h$. 

\begin{figure}[t!]
    \centering
    \begin{subfigure}[b]{0.48\textwidth}
        \centering
        \includegraphics[width=\textwidth]{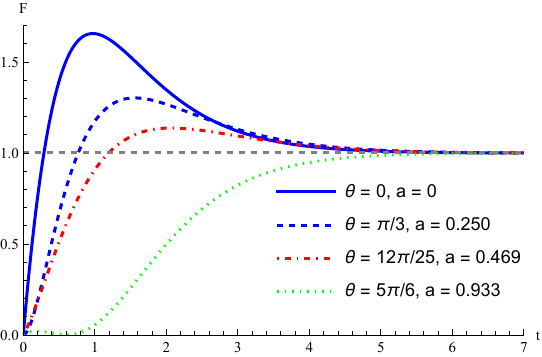}
        \caption{Cold reservoir, $\beta_c = 0.0334$.}
        \label{fig:FI_cold_coh}
    \end{subfigure}
    \hfill
    \begin{subfigure}[b]{0.48\textwidth}
        \centering
        \includegraphics[width=\textwidth]{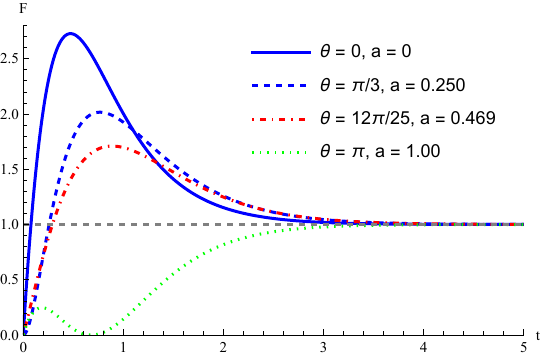}
        \caption{Hot reservoir, $\beta_h \approx 0.020$.}
        \label{fig:FI_hot_coh}
    \end{subfigure}
    \caption{QFI over time by initializing the qubit thermometer in the density operator \eqref{eq:qubit_state_parametrization} with $r=1$, $\phi=0$ and $a=\sin^2(\theta/2)$. We also set $\omega_{12}=5$ and $\widetilde{\tau}=0.05$. Moreover, in panel (a), we consider $\theta=0,\pi/3,12\pi/25,5\pi/6$ and $n_{12}=5.5$ that means $\beta_c = 0.0334$. Conversely, in panel (b), we take $\theta=0,\pi/3,12\pi/25,\pi$ and $n_{12}=9.5$ leading to $\beta_h \approx 0.020$.}
    \label{fig:QFI_experiment}
\end{figure}

In Fig.~\ref{fig:QFI_experiment} we plot the QFI $F(\rho_{\beta}(t),\beta)$ of Eq.~\eqref{eq:QFI_t} as a function of time, for $\theta=0,\pi/3,12\pi/25,5\pi/6$ [panel (a)] and $\theta=0,\pi/3,12\pi/25,\pi$ [panel (b)]. The two panels of Fig.~\ref{fig:QFI_experiment} differ for the value of the inverse temperature: $\beta_c \approx 0.033$ in panel (a) and $\beta_h \approx 0.020$ in panel (b). In both cases, the greatest value of the QFI is obtained by setting $\theta=0$ (i.e., $a=0$) in the initial transient. For the experiments in \cite{MancinoPRL2017}, $\theta=0$ corresponds to initialize the single photons encoding the qubit thermometer in the horizontal or vertical polarization state. Interestingly, there are time intervals (before the qubit is fully thermalized) where the QFI is not maximized by initializing the qubit thermometer in the ground state of $H_0$, i.e., by setting $\theta=0$. 

\begin{figure}[t!]
    \centering
    \begin{subfigure}[b]{0.45\textwidth}
        \centering
        \includegraphics[width=\textwidth]{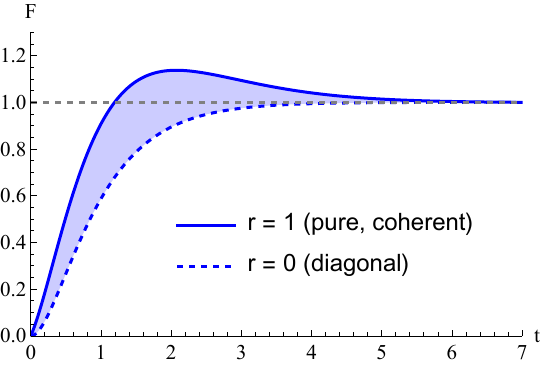}
        \caption{Cold reservoir, $\beta_c = 0.0334$.}
        \label{fig:FI_cold_coh}
    \end{subfigure}
    \hfill
    \begin{subfigure}[b]{0.45\textwidth}
        \centering
        \includegraphics[width=\textwidth]{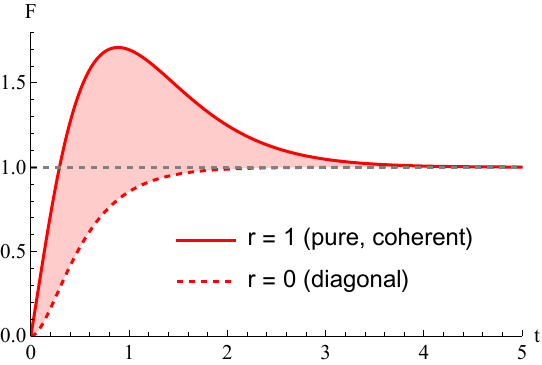}
        \caption{Hot reservoir, $\beta_h \approx 0.020$.}
        \label{fig:FI_hot_coh}
    \end{subfigure}
    \caption{Role of coherence in the initial state. QFI over time in the realistic scenario realized in \cite{MancinoPRL2017}. The qubit thermometer is initialized in the density operator \eqref{eq:qubit_state_parametrization} with $r=1$ or $r=0$. As in Fig.~\ref{fig:QFI_experiment}, we have in panel (a) $n_{12}=5.5$ and $\beta_c = 0.0334$, while in panel (b) $n_{12}=9.5$ and $\beta_h \approx 0.020$. In both panels we set $\phi=0$, $\theta=12\pi/25$ thus $a=0.469$, $\omega_{12}=5$ and $\widetilde{\tau}=0.05$.}
    \label{fig:QFI_experiment_2}
\end{figure}

We conclude by showing in Fig.~\ref{fig:QFI_experiment_2} the advantage entailed by the presence of quantum coherence in the initial state of the thermometer. In agreement with Fig.~\ref{fig:FI_coh_comparison}, we can observe an apparent advantage for both the cold and hot temperatures considered in \cite{MancinoPRL2017}. Moreover, it is also evident that, as long as $r=0$, the three time-behaviours of the QFI in the regions $\mathcal C$, $\mathcal H$ and $\mathcal I$ described in Fig.~\eqref{fig:FI2_phase_space} are recovered. This is not necessarily true by initializing the qubit thermometer in a pure state ($r=1$). 

For the sake of an interpretation, the presence of quantum coherence in the thermometer's state entails a higher purity of the state itself. The latter can be mapped unitarily to a diagonal density operator whose elements can be linked to a smaller effective temperature. Thus, from this point of view, the increase in the QFI could be seen as a cooling effect on the thermometer's state due to quantum coherence, provided the amount of purity is kept the same (a unitary mapping is indeed assumed).

Based on these results, further experimental tests are foreseeable, provided the availability of an estimation method/algorithm that returns the estimated inverse temperature using the density operator of the qubit thermometer (to be got via tomography) in the time interval where the QFI is maximized. In this way, one could explore if the presence of quantum coherence in the initial states of the thermometer shows up even in a smaller estimation error in a given nonequilibrium regime.   

\subsection*{Acknowledgments}

GF acknowledges support from FCT -- Funda\c{c}\~{a}o para a Ci\^{e}ncia e a Tecnologia (Portugal) through scholarship SFRH/BD/145572/2019. 
MP and SG wish to acknowledge financial support from the PRIN project 2022FEXLYB Quantum Reservoir Computing (QuReCo). SG also acknowledges the PNRR MUR project PE0000023-NQSTI funded by the European Union--Next Generation EU.
YO thanks the support from FCT through project UIDB/04540/2020.

\appendix
\setcounter{equation}{0}

\section{On the eigenvalues of the transition matrix $A_{\beta}$}
\label{app:Aeig}

In this Appendix we show that the transition matrix $A_\beta$, from Eq.~\eqref{eq:A}, has a single null eigenvalue, whereas the remaining $N-1$ eigenvalues have negative real part. Furthermore we show that the eigenvector with null eigenvalue is the thermal population $\boldsymbol{\pi}(\beta)$.\\

In continuous-time Markov chain theory, the state of a Markov chain at time $t$ is described by a probability vector $\mathbf p(t)$, where $p_i(t)$ is the probability of the chain being in state $i$. A Markov process is characterized by a transition rate matrix $A$, where the off-diagonal entries $a_{ij}$ ($i\neq j$) denote the instantaneous transition rate from state $j$ to state $i$. The evolution of the process is given by the differential equation $\dot{ \mathbf p}(t) = A\, {\mathbf p}(t)$ or the equivalent integral equation $\mathbf p(t)= e^{A(t-t_0)}\mathbf p(t_0)$.

For the thermalization model in Sec.~\ref{sec:model}, the transition matrix $A_\beta$, which describes the evolution of the population terms, exhibits the following properties:
\begin{enumerate}
    \item[1.] $A_\beta=[a_{ij}]_{ij}$ is a real square matrix, that is, $A_\beta \in \mathbb R^{N\times N}$.
    \item[2.] The off-diagonal elements are positive, that is, $a_{ij} > 0$ $\forall_{i\neq j}$.
    \item[3.] The columns sum to zero, that is, $\sum_{i=1}^N a_{ij} = 0$ $\forall j$, as a direct consequence of probability conservation, see Eqs.~\eqref{eq:rho_dot_ii_N} and \eqref{eq:a_ii}.
\end{enumerate}
\noindent To prove that the transition matrix $A_\beta$ has only one null eigenvalue and remaining $N-1$ eigenvalues with a negative real part, we first show that all the eigenvalues have non-positive real part, and then demonstrate that there is only one null eigenvalue.

\subsection{The real part of the eigenvalues of $A_\beta$ is non-positive}
To study the eigenvalues of $A_\beta$, we introduce the following theorem:
\begin{thm}[Gershgorin circle theorem \cite{Gershgorin1931}]
Let $A$ be a complex $ N\times N$ matrix. Define $R_j$ as the sum of the absolute values of the non-diagonal entries in the $j^{\text th}$ column: $R_j = \sum\limits_{\substack{i=1 \\ i\neq j}}^N |a_{ij}|.$

Let $D(a_{jj}, R_j)=\{z \in \mathbb C: |z-a_{jj}| \leq R_j \}$ be the closed disc centered at $a_{jj}$ with radius $R_j$. Such a disc is called a Gershgorin disc.
Then each eigenvalue of $A$ lies within at least one of the Gershgorin discs $D(a_{jj}, R_j)$.
\end{thm}

Applying the Gershgorin circle theorem to our matrix $A_\beta$, together with properties 2. and 3. above, each eigenvalue $\nu$ must satisfy, for at least one $j \in \{1,\dots,N\}$, the inequality:
\begin{align}
    |\nu - a_{jj}| \le \sum_{\substack{i=1 \\ i\neq j}}^N |a_{ij}| = - a_{jj} = |a_{jj}|,
\end{align}
which implies that any eigenvalue of $A_\beta$ has non-positive real part.

\subsection{$A_\beta$ has a single null eigenvalue}

The existence of a null eigenvalue of $A_\beta$ follows immediately from the fact that the columns of the matrix sum up to $0$. To show this, construct an auxiliary $N\times N$ matrix $B$, equal to $A_\beta$, but with the last row zeroed. Since row operations do not change the nullity of a matrix, $B$ has the same number of null eigenvalues as $A_\beta$.
\begin{align}
B = \begin{bmatrix}
a_{11} & \cdots & a_{1(N-1)} & a_{1N} \\
\vdots & \ddots & \vdots & \vdots \\
a_{1(N-1)} & \cdots & a_{(N-1)(N-1)} & a_{(N-1)(N-1)} \\
0 & \cdots & 0 & 0
\end{bmatrix}
\end{align}
Construct also a second auxiliary $(N-1) \times (N-1)$ matrix $C$ by removing the last row and column of $B$. Note that each eigenvalue $\lambda$ of $C$ is also an eigenvalue of $B$ since you can construct eigenvectors of $B$ by padding the eigenvectors of $C$ with a $0$:
\begin{align*}
&\text{Let} \quad \mathbf v=[v_1\ \cdots\ v_{N-1}] \quad \text{ such that} \quad C\mathbf v = \lambda \mathbf v; \\
&\text{denote} \quad \mathbf w=[v_1\ \cdots\ v_{N-1} \ 0] \quad \Longrightarrow \quad B\mathbf w = \lambda \mathbf w.
\end{align*}
Use once again the Gershgorin circle theorem to find that each eigenvalue of $C$ has strictly negative real part:
\begin{align}
|\lambda - a_{jj}| \leq \sum_{\substack{i=1 \\ i\neq j}}^{N-1} |a_{ij}| = |a_{jj}| - a_{Nj} < |a_{jj}|.
\end{align}
$C$ has no null eigenvalues, thus $B$ and, consequently, $A_\beta$ have only one null eigenvalue, which completes the proof.

\subsection{Proof of $A_\beta \, \boldsymbol{\pi}=0$}\label{app:A_th}

The verification of the thermal distribution $\boldsymbol{\pi}(\beta)$ as the null eigenvector of $A_\beta$ can be checked via direct substitution.
For this purpose, we rewrite the entries $a_{ij}$ of $A_\beta$, Eqs.~(\ref{eq:a_ij}) and (\ref{eq:a_ii}), as a function of the thermal population $\pi_k = e^{-\beta\epsilon_k}/Z_{\beta}$ with $Z_{\beta} = \sum_{k}e^{-\beta\epsilon_k}$:
\begin{align}
    n_{ij} &= \bigl(e^{\beta \omega_{ij}}-1\bigr)^{-1} = \biggl(\frac{\pi_i}{\pi_j}-1\biggr)^{-1} = \frac{\pi_j}{\pi_i-\pi_j},\\[2 ex]
    a_{ij}=\Gamma_{ij} &= \gamma\begin{cases}
    n_{ij}+1 &= \pi_i/(\pi_i-\pi_j) \quad \text{for} \quad i<j \\[2ex]
    n_{ji} &= \pi_i/(\pi_j-\pi_i) \quad \text{for} \quad i>j \,.
    \end{cases}\label{eq:etas}
\end{align}
Notice that the off-diagonal elements of the transition matrix verify the detailed balance equation $a_{ij} \pi_j = a_{ji}\pi_i$ defined over the terms of the thermal distribution $\boldsymbol{\pi}$. In fact,
\begin{equation}\label{eq:detailed_balance}
    \frac{a_{ij}}{a_{ji}} = \frac{ \frac{\pi_i}{\pi_i-\pi_j} }{ \frac{\pi_j}{\pi_i-\pi_j} } =  \frac{\pi_i}{\pi_j} \quad
    \Longrightarrow \quad a_{ij} \pi_j = a_{ji}\pi_i \,. 
\end{equation}
Finally, using the detailed balance equation (\ref{eq:detailed_balance}) together with the fact that the columns of $A_\beta$ sum up to $0$ (property 3.~at the beginning of this Appendix), we can prove that $\boldsymbol{\pi}$ is the eigenvector with null eigenvalue:
\begin{align}
    a_{ij} \pi_j = a_{ji}\pi_i\Rightarrow A_\beta \boldsymbol{\pi} = \begin{bmatrix} \sum_k a_{ik}\pi_k \end{bmatrix}_i =  \begin{bmatrix} \pi_i\sum_k a_{ki}\end{bmatrix}_i = {\bf 0} \,.
\end{align}

\section{Derivatives of thermometer's state w.r.t.~the inverse temperature}

\subsection{Derivation of Eqs.~\eqref{eq:partial_p} and \eqref{eq:partial_rho12}}\label{app:partial_deriv}

We aim to compute the derivatives of the population and coherence elements of a qubit thermometer, i.e.,
\begin{align}
    \mathbf p(t,\beta,a) &= (1- e^{\lambda t})\boldsymbol{\pi} + e^{\lambda t}\mathbf p_0(a) \\
    \rho_{12}(t,\beta) &= e^{\lambda t/2} e^{\text{i}\,\omega_{12}t} \rho_{12}(0),
\end{align}
with respect to $\beta$. 

For this purpose, we start from calculating the derivative of the thermal population elements $\boldsymbol{\pi}$:
\begin{align}
    \partial_\beta\left(\pi_2\right) &= (\epsilon_1(1-\pi_2)+\epsilon_2\pi_2-\epsilon_2)\pi_2 = (1-\pi_2)\pi_2\,\omega_{12}\,,\\
    \partial_\beta\left(\pi_1\right) &= -\partial_\beta\left(\pi_2\right).
\end{align}
Moreover, the derivative of the eigenvalue $\lambda=(\pi_1-\pi_2)^{-1}$ w.r.t.~$\beta$ is
\begin{align}
    \partial_\beta\left(\lambda\right) &= \partial_\beta\left( (\pi_2-\pi_1)^{-1}\right) = 2\lambda^2\partial_\beta\left(\pi_1\right) = -2\lambda^2\partial_\beta\left(\pi_2\right).
\end{align}

Accordingly, the explicit expression for $\partial_\beta\left(\mathbf p(t,\beta,a)\right)$ reads as:
\begin{align}
    \partial_\beta\left(\mathbf p(t,\beta,a)\right) &= \partial_\beta\left(\boldsymbol{\pi}\right) - e^{\lambda t}\partial_\beta\left(\boldsymbol{\pi}\right) -t\partial_\beta\left(\lambda\right)e^{\lambda t}(\boldsymbol{\pi}-\mathbf p_0(a)) = \nonumber\\
    &= \begin{bmatrix}
        \partial_\beta\left(\pi_1\right)\bigl(1-e^{\lambda t}-2t\lambda^2e^{\lambda t}(\pi_1 - 1+a)\bigr)\\
        \partial_\beta\left(\pi_2\right)\bigl(1-e^{\lambda t}+2t\lambda^2e^{\lambda t}(\pi_2 - a)\bigr)
    \end{bmatrix} = \nonumber\\
    &= (1-\pi_2)\pi_2\,\omega_{12}\,\bigl(1-e^{\lambda t}+2t\lambda^2e^{\lambda t}(\pi_2 - a)\bigr) \begin{bmatrix}
        -1 \\ 1
    \end{bmatrix} = \nonumber\\
    &= (1-\pi_2)\pi_2\,\omega_{12}\,\delta(t,\beta,a)\, \begin{bmatrix}
        -1 \\ 1
    \end{bmatrix},
\end{align}
where $\delta(t,\beta,a) =  1-e^{\lambda t} + 2t\lambda^2 e^{\lambda t}(\pi_2-a)$.

On the other hand, the explicit expression for $\partial_\beta\left(\rho_{12}(t,\beta)\right)$ is
\begin{align}
    \partial_\beta\left(\rho_{12}(t,\beta)\right) &= \frac12 t\,\partial_\beta\left(\lambda\right) e^{\lambda t/2} e^{\text{i}\,\omega_{12}t} \rho_{12}(0) = \nonumber\\
    &= \alpha(t,\beta) \, \rho_{12}(t,\beta),
\end{align}
where $\alpha(t,\beta) = -(1-\pi_2)\pi_2\,\omega_{12} \lambda^2 t$.

\subsection{Comment on the difference between partial and total derivative}\label{appendix_sec:partial_vs_total_deriv}

Let us consider a qubit thermometer, and the thermalization process that transforms a given input distribution $\mathbf p_0(a)$ to an output distribution $\mathbf p(t,\beta,a):\mathbf p_0(a) \rightarrow \mathbf p(t,\beta,a)$.
Here, we will take into account the dependence on $\beta$ whenever necessary, being it important for this discussion.
In accordance with the thermalization dynamics in Sec.~\ref{sec:model}, the probability $\mathbf p(t,\beta,a)$ evolves as 
\begin{equation}
    \mathbf p(t,\beta,a) = \boldsymbol{\pi}(\beta)- e^{\lambda(\beta)t}\bigl(\boldsymbol{\pi}(\beta)-\mathbf p_0(a)\bigr) = \bigl(1- e^{\lambda(\beta)t}\bigr)\boldsymbol{\pi}(\beta)+ e^{\lambda(\beta)t}\mathbf p_0(a) \,,    
\end{equation}
where $\lambda(\beta) = \gamma(\pi_2-\pi_1)^{-1} <0$ $\forall \beta$ guarantees the convergence of the system dynamics. In fact, as the time increases, the distribution $\mathbf p(t,\beta,a)$ is converging asymptotically to the thermal distribution $\boldsymbol{\pi}(\beta) = \lim_{t\rightarrow \infty} \mathbf p(t,\beta,a)$. Such an asymptotic convergence occurs irrespectively of the initial distribution $\mathbf p_0(a)$. Moreover, the derivative of $\mathbf p(t,\beta,a)$ w.r.t.~$\beta$ is given by:
\begin{align}
    \partial_\beta\left(\mathbf p(t,\beta,a)\right) &= \partial_\beta( \boldsymbol{\pi}(\beta) ) - e^{\lambda(\beta)t}\partial_\beta( \boldsymbol{\pi}(\beta) ) - t\partial_\beta( \lambda(\beta) ) e^{\lambda(\beta)t}\bigl(\boldsymbol{\pi}(\beta)-\mathbf p_0(a)\bigr) \,.
\end{align}

We now analyze the case where, as input to the process, we set an initial distribution that is equal to the asymptotic distribution, i.e., $\mathbf p_0(a) = \boldsymbol{\pi}(\beta)$. The distributions $\mathbf p_0(a)$ and $\boldsymbol{\pi}(\beta)$ are parameterized as
\begin{equation}
\mathbf p_0(a) =\begin{bmatrix}1-a\\ a \end{bmatrix}
\quad \textrm{and} \quad
\boldsymbol{\pi}(\beta)=\begin{bmatrix}1-\pi_2(\beta)\\ \pi_2(\beta)\end{bmatrix}   .
\end{equation}
If the initial distribution $\mathbf p_0(a)$ is equal to the thermal distribution $\boldsymbol{\pi}(\beta)$, that is $a=\pi_2(\beta)$, then $\mathbf p\bigl(t,\beta,\pi_2(\beta)\bigr)$ is constant over time, as $\boldsymbol{\pi}(\beta)$ is a fixed point of the thermalization process. In particular,
\begin{equation}\label{eq:p_beta}
    \mathbf p\bigl(t,\beta,\pi_2(\beta)\bigr) = \boldsymbol{\pi}(\beta)
\end{equation}
and
\begin{equation}\label{eq:dp_beta}
    \partial_\beta\left(\mathbf p(t,\beta,a)\right)\Big|_{a=\pi_2(\beta)}
    =  \partial_\beta(\boldsymbol{\pi}(\beta))- e^{\lambda(\beta)t}\partial_\beta(\boldsymbol{\pi}(\beta)) 
    = \left( 1 - e^{\lambda(\beta)t} \right)\partial_\beta(\boldsymbol{\pi}(\beta))\,. 
\end{equation}
It is worth noting that the right-hand-side of Eq.~(\ref{eq:dp_beta}) is different from the partial derivative of the right-hand-side of Eq.~(\ref{eq:p_beta}) w.r.t.~$\beta$. Formally, 
\begin{equation}\label{eq:difference_partial_total_der}
\partial_\beta\left(\mathbf p(t,\beta,a)\right)\Big|_{a=\pi_2(\beta)} \neq \partial_\beta\left(\mathbf p\bigl(t,\beta,\pi_2(\beta)\bigr)\right) = \partial_\beta(\boldsymbol{\pi}(\beta))\,.
\end{equation}
The discrepancy in Eq.~\eqref{eq:difference_partial_total_der} can be resolved by considering how the partial derivative $\partial_\beta = \partial/\partial\beta$ differs from the total derivative $d_\beta = d/d\beta$. Let us argue about that. From the one hand, the partial derivative of $p(t,\beta,a)$ w.r.t.~$\beta$ leads to 
\begin{equation}
     \partial_\beta(\mathbf p(t,\beta,a)) = \partial_\beta(\boldsymbol{\pi}(\beta)) - e^{\lambda(\beta)t}\partial_\beta(\boldsymbol{\pi}(\beta))-t\partial_\beta(\lambda(\beta)) e^{\lambda(\beta)t}\bigl(\boldsymbol{\pi}(\beta)-\mathbf p_0(a)\bigr)\,,
\end{equation}
whereby
\begin{equation}
    \partial_\beta\left(\mathbf p(t,\beta,a)\right)\Big|_{a=\pi_2(\beta)} = \partial_\beta(\boldsymbol{\pi}(\beta)) - e^{\lambda(\beta)t}\partial_\beta(\boldsymbol{\pi}(\beta))\,.
\end{equation}
On the other hand, instead, the total derivative is the correct operation that leads to the equality
\begin{equation}
    d_\beta\left(\mathbf p\left(t,\beta,a(\beta)=\pi_2(\beta)\right)\right) = \partial_\beta\left(\mathbf p\bigl(t,\beta,\pi_2(\beta)\bigr)\right) = \partial_\beta(\boldsymbol{\pi}(\beta))\,.
\end{equation}
In fact, it holds that
\begin{align}
    d_\beta\left(\mathbf p\left(t,\beta,a(\beta)=\pi_2(\beta)\right)\right) &= \partial_\beta\left(\mathbf p(t,\beta,a)\right)\Big|_{a=\pi_2(\beta)}
    + \partial_a\left( \mathbf p(t,\beta,a(\beta))\right)
    ~d_{\beta}a(\beta)\\
    &=\partial_\beta(\boldsymbol{\pi}(\beta))- e^{\lambda(\beta)t}\partial_\beta(\boldsymbol{\pi}(\beta))+
    \partial_a \left( e^{\lambda(\beta)t}\begin{bmatrix}1-a\\ a \end{bmatrix}\right)\partial_\beta(\pi_2(\beta))\nonumber\\
    &=\partial_\beta(\boldsymbol{\pi}(\beta))- e^{\lambda(\beta)t}\partial_\beta(\boldsymbol{\pi}(\beta))+e^{\lambda(\beta)t}\partial_\beta( \pi_2(\beta))\begin{bmatrix}-1\\ 1\end{bmatrix}\nonumber\\
     &=\partial_\beta(\boldsymbol{\pi}(\beta))- e^{\lambda(\beta)t}\partial_\beta(\boldsymbol{\pi}(\beta))+e^{\lambda(\beta)t}\partial_\beta(\boldsymbol{\pi}(\beta))\nonumber\\
     &=\partial_\beta(\boldsymbol{\pi}(\beta))
\end{align}
with $\partial_a \equiv \partial/\partial a$.

\subsection{Derivation of Eq.~\eqref{eq:partial_rho_beta_main}}\label{app_sec:QFI_thermal_state}

In this Appendix we show the derivation of 
\begin{equation}\label{eq:app_derv_thermal}
    \partial_{\widetilde{\beta}}( \rho_{\widetilde{\beta}} ) = \sum_{j=1}^{N}\partial_{\widetilde{\beta}}\left( \pi_j(\widetilde{\beta}) \right) |\epsilon_j\rangle\!\langle \epsilon_j|\,, 
\end{equation}
where $\widetilde{\beta}$ is the inverse temperature of an initial thermal state, which should not be confused with the inverse temperature $\beta$ of the thermal bath.
As explained in the main text, the computation of Eq.~\eqref{eq:app_derv_thermal} is the requisite to achieve the analytical expression of the QFI of the thermal state $\rho_{\widetilde{\beta}}$ in \eqref{eq:rho_beta_in}. We recall that the latter is given by $F(\rho_{\widetilde{\beta}},\widetilde{\beta}) = {\rm Var}(H)$, where ${\rm Var}(H)_{\rho_{\widetilde{\beta}}}$ is the variance of $H$ computed w.r.t.~the thermal state $\rho_{\widetilde{\beta}}$.

To derive $\partial_{\widetilde{\beta}}( \rho_{\widetilde{\beta}} )$, we compute the derivative of the thermal probabilities $\pi_{j}(\widetilde{\beta})=e^{-\widetilde{\beta}\epsilon_j} / Z_{\widetilde{\beta}}$ with respect  $\widetilde{\beta}$. We recall that $Z_{\widetilde{\beta}} = \sum_{j=1}^{N}e^{-\widetilde{\beta}\epsilon_j}$. Thus,  
\begin{equation}\label{eq:dpij_dbeta}
    \partial_{\widetilde{\beta}}\left( \pi_{j}(\widetilde{\beta}) \right) = \frac{1}{Z^{2}(\widetilde{\beta})}
    \left(\partial_{\widetilde{\beta}}( e^{-\widetilde{\beta}\epsilon_{j}}) Z(\widetilde{\beta}) - e^{-\widetilde{\beta}\epsilon_j}\partial_{\widetilde{\beta}}( Z(\widetilde{\beta}) ) \right).
\end{equation}
Then, we determine $\partial_{\widetilde{\beta}}( Z(\beta) )$. To do this, it is worth writing the explicit expression of the average of $H$ w.r.t.~$\rho_{\widetilde{\beta}}$:
\begin{equation}
    \langle H\rangle_{\rho_{\widetilde{\beta}}} = {\rm Tr}\left[ \rho_{\widetilde{\beta}}H \right] = \sum_{j=1}^N \epsilon_j \pi_j(\widetilde{\beta}) = \frac{1}{Z(\widetilde{\beta})}\sum_{j=1}^N \epsilon_j e^{-\widetilde{\beta}\epsilon_j} = - \frac{1}{Z(\widetilde{\beta})} \partial_{\widetilde{\beta}}Z(\widetilde{\beta})\,.
\end{equation}
Therefore, 
\begin{equation}\label{eq:dZ_dbeta}
    \partial_{\widetilde{\beta}} Z(\widetilde{\beta}) = - \langle H\rangle_{\rho_{\widetilde{\beta}}} Z(\widetilde{\beta}).
\end{equation}
As a result, substituting \eqref{eq:dZ_dbeta} into \eqref{eq:dpij_dbeta},
\begin{equation}\label{eq:partial_pi}
    \partial_{\widetilde{\beta}}\left( \pi_{j}(\widetilde{\beta}) \right) = \left( \langle H\rangle_{\rho_{\widetilde{\beta}}} - \epsilon_j \right)\pi_j(\widetilde{\beta}),
\end{equation}
so that
\begin{equation}
    \partial_{\widetilde{\beta}}\rho_{\widetilde{\beta}} = \sum_{j=1}^N \left( \langle H\rangle_{\rho_{\widetilde{\beta}}} - \epsilon_j \right)\pi_j(\widetilde{\beta})|\epsilon_j\rangle\!\langle \epsilon_j|\,. 
\end{equation}

\bibliography{biblio}

\end{document}